\documentclass[iop,apj,tighten]{emulateapj}
\pdfoutput=1 
\usepackage{apjfonts} 
\usepackage{amsmath,amstext}
\usepackage[breaklinks,colorlinks,citecolor=blue,linkcolor=magenta]{hyperref} 

\shorttitle{CAFE-PakalMPI}
\shortauthors{Gonz\'alez-Avil\'es \& De la Luz.}

\begin{document}

\title{CAFE-PakalMPI: a new model to study the solar chromosphere in the NLTE approximation}

\author{J.J. Gonz\'alez-Avil\'es \altaffilmark{1} and V. De la Luz \altaffilmark{1,2}}
\affil{$^1$ Instituto de Geof\'isica, Unidad Michoac\'an, Universidad Nacional Aut\'onoma de M\'exico, Morelia, Michoac\'an, M\'exico.\\
$^2$ Conacyt, SCiESMEX, LANCE, Instituto de Geof\'isica, Unidad Michoac\'an, Universidad Nacional Aut\'onoma de M\'exico, Morelia, Michoac\'an, M\'exico.}

\begin{abstract}
We present a new numerical model called CAFE-PakalMPI with the capability to solve the equations of classical magnetohydrodynamic (MHD) and to obtain the multispecies whose ionization states are calculated through statistical equilibrium, using the approximation of non-local thermodynamic equilibrium (NLTE) in three dimensions with the multiprocessor environment. For this, we couple the Newtonian CAFE MHD code introduced in \cite{CAFE_code} with PakalMPI presented in \cite{De_la_Luz_et_al_2010}. In this model, Newtonian CAFE solves the equations of ideal MHD under the effects of magnetic resistivity and heat transfer considering a fully ionized plasma. On the other hand, PakalMPI calculates the density of ionization states using the NLTE approximation for Hydrogen, electronic densities and H-, the other species are computed by the classical local thermodynamic equilibrium (LTE) approximation. The main purpose of the model focuses on the analysis of solar phenomena within the chromospheric region. As an application of the model, we study the stability of the C7 equilibrium atmospheric model with a constant magnetic field in a 3D environment. According to the results of the test, the C7 model remains stable in the low chromosphere, while in the range $z\in$[1.5,2.1] Mm we can observe the propagation of a wave produced by the changes in the ionization rate of H. 
\end{abstract}


\section{Introduction}

The solar atmosphere is the most dynamic layer of the Sun, here the fully or partially ionized plasma interacts with magnetic fields, this interaction can produce different kind of waves, jets, and other phenomena. In particular, the chromosphere-transition region is a zone of the solar atmosphere, where the plasma is not fully ionized. Here the plasma goes from nearly neutral to almost completely ionized, the temperature increases from a few thousand Kelvin to a million Kelvin, and the dynamics goes from being dominated by the gas pressure to being dominated by the magnetic pressure \citep{Golding_et_al_2016}. The most abundant element in the Sun is the Hydrogen, its ionization does not obey local thermodynamic equilibrium (LTE) or instantaneous statistical equilibrium because the timescale is long compared with important hydrodynamical timescales, especially of magneto-acoustic shocks \citep{Leenaarts_et_al_2007}. Usually, Hydrogen ionization is calculated from the condition of statistical equilibrium, that is, the equality of the ionization and recombination rates for the local temperature, density, and radiation. Statistical equilibrium assumes infinitely fast rates and an adjustment to the local thermodynamic and radiation state \citep{Carlsson&Stein_2002}. An account of the non-equilibrium ionization states involves solving a set of rate equations for the atomic population densities. The transition rate coefficients involve frequency integrals over the intensity, therefore a general description must necessarily take into account the radiative transfer, which in 1D is computationally doable, but in 2D or 3D a detailed treatment of radiative transfer is challenging.  
 
 The reasons mentioned above have motivated the development of numerical tools capable to solve the full MHD equations coupled with the calculation of non-equilibrium ionization states including radiative processes mimicking the most realistic conditions in the solar atmosphere \citep{Hansteen_et_al_2007,Leenaarts_et_al_2007}. For this purpose a number of codes have been developed, for instance, the {\it Bifrost} code \citep{Gudiksen_et_al_2011}, designed to simulate stellar atmospheres from the convection zone to the corona, solves the MHD equations considering the effects of thermal conduction and full radiative transfer in the LTE and NLTE approximations. {\it MURaM} code \citep{Vogler_2004,Rempel_et_al_2009}, developed for applications in the solar convection zone and photosphere, solves the non-ideal MHD equations considering the non-local and non-grey radiative transfer and takes into accounts the effects of partial ionization. {\it CO$^{5}$BOLD} code \citep{Freytag_et_al_2012}, designed for realistic simulations that take into account the detailed microphysics of the solar or stellar surfaces layers, solves the hydrodynamic and MHD equations including gravity, radiative energy exchange and Hydrogen ionization. {\it ANTARES} code \citep{Muthsam_et_al_2010}, developed for simulations in stellar hydrodynamics with radiative transfer and realistic microphysics in 1D, 2D and 3D. 

Unlike the complexity of the codes mentioned above, in this paper, we present a model that do not solve the equations of full radiative MHD, neither it solves the rate equations for the atomic populations densities coupled with the MHD equations, instead, our model first solves the equations of classical MHD under the effects of magnetic resistivity and heat transfer of a fully ionized plasma in order to obtain post-processing the dynamics of multispecies whose ionization states are calculated through statistical equilibrium, using the LTE and NLTE approximations. The numerical model consists of two parts: i) Newtonian CAFE solves the MHD equations for a fully ionized gas to obtain the total mass density and temperature for each time-step, ii) PakalMPI uses the total mass density and temperature calculated with CAFE to obtain the ionization states using the NLTE approximation for Hydrogen, electronic densities and H-, the other species are computed by classical LTE approximation each time-step.   

The overview of Newtonian CAFE code, Pakal code and the coupling of the CAFE-PakalMPI model is described in Section \ref{CAFE_pakal_model}. The study of the stability of the C7 model and the convergence are presented in Section \ref{results}. In Section \ref{conclusions}, we present our conclusions and final comments. 

\section{CAFE-PakalMPI model}
\label{CAFE_pakal_model}

\subsection{Newtonian CAFE code}

Newtonian CAFE is a code originally designed to solve the equations of classical ideal magnetohydrodynamics (MHD) in three dimensions, submitted to a constant gravitational field \citep{CAFE_code}. The initial purpose of the code was the analysis of solar phenomena within the photosphere-corona region. Newtonian CAFE has been recently improved to solve the equations of MHD under the effects of magnetic resistivity and heat transfer \citep{CAFE_Q}, with the main objective to study the formation of jets in the solar atmosphere and the propagation of Coronal Mass Ejections (CMEs) in the interplanetary medium, as well as to enhance the state of the art simulations related to the formation of different scale jets in the solar atmosphere \citep{Martinez-Sykora_et_al_2017,De_Pontieu_et_al_2017,van_der_Voort_et_al_2017}. The numerical algorithms implemented in the code are based on high-resolution shock-capturing methods \citep{LeVeque_1992}, using the Harten-Lax-van Leer-Einfeldt (HLLE), Harten-Lax-van Leer-Contact (HLLC) and the Harten-Lax-van Leer-Discontinuities (HLLD) flux formulas \citep{Harten_Lax_Van_Leer_1983,Einfeldt_1988,Li_2005,Miyoshi&Kusano_2005} combined with MINMOD, MC and WENO5 reconstructors \citep{Harten_et_al_1997,Titarev&Toro_2004,Radice&Rezzolla_2012}. The divergence free magnetic field constraint is controlled using the extended generalized Lagrange Multipliers Method (EGLM) \citep{Dedner_et_al_2002} and the Flux Constrained Transport (Flux-CT) method \citep{Evans&Hawley_1988,Balsara_2001}. The code has been tested with 1D and 2D standard tests to demonstrate the quality of the numerical methods implemented \citep{CAFE_code}. The code has been applied to study the influence of Alfv\'en waves in the process of heating the quiet solar corona \citep{Gonzalez-Aviles&Guzman_2015}, the formation of jets with features of Type II spicules and cool coronal jets as a result of magnetic reconnection \citep{Gonzalez_Aviles_et_al_2017}. Recently, it has been used to study that 3D magnetic reconnection may be responsible of the formation of jets with characteristics of Type II spicules \citep{Gonzalez_Aviles_et_al_2018a,Gonzalez_Aviles_et_al_2018b}.

\subsection{PakalMPI}

PakalMPI was designed to solve the radiative transfer equation in 3D using a non-local thermodynamic equilibrium (NLTE) approximation \citep{De_la_Luz_et_al_2010,De_la_Luz_et_al_2011}. The code receive as input the profiles of Hydrogen, Temperature, a set of abundances, and the departure coefficients for Hydrogen that represents an stratified hydrostatic atmosphere. To solve the system, the model have two main steps: i) compute the physical conditions for all the layers of the atmosphere and ii) solve the radiative transfer equation in a 3D geometry. The physical conditions includes the NLTE computations for the following species: HI, HII, H-, and $n_e$ using an approximation described in \cite{De_la_Luz_et_al_2011}. The contribution of the other atoms are computed in the classical local thermodynamic equilibrium using the Saha Equation. With the abundance profiles, PakalMPI solves the radiative transfer equation in a set of 3D ray paths using opacities focused in the sub-millimeter emission. PakalMPI has been used in the study the emission in the solar temperature minimum region \citep{De_la_Luz_et_al_2013}, the quiet sun emission in the continuum \citep{De_la_Luz_et_al_2016}, solar flares \citep{2015SoPh..290.2809T}, and chromospheric emission in solar like stars \citep{2015A&A...573L...4L}. The design of PakalMPI allows us to integrate the results from CAFE as data input for PakalMPI. However, the computations for abundances was not parallelized, for this reason in the CAFE-PakalMPI model we developed an efficient parallelization method to optimize the computations in 3D.

\subsection{CAFE-PakalMPI model}

CAFE-PakalMPI model is the result of coupling the Newtonian CAFE with PakalMPI. In this new model, Newtonian CAFE solves the equations of classical ideal magnetohydrodynamic in three dimensions, submitted to a constant gravitational field to obtain the profiles of total mass density $\rho$ and temperature $T$, with these profiles PakalMPI calculates the multispecies whose ionization states are obtained using the LTE and NLTE approximations. The 3D data of total mass density $\rho$ and temperature $T$ computed by CAFE for each time step are taking as a set of input profiles for the start of computations of PakalMPI. 
 
Following \cite{De_la_Luz_et_al_2011}, we explain the method to calculate the density profiles for the ionization states in our model. Firstly, PakalMPI needs the Hydrogen density, the temperature, and the departure coefficients for the atmospheric model. In our CAFE-PakalMPI model, as a first approximation we take the total mass density $\rho$  calculated with CAFE as the total Hydrogen density, and the temperature is calculated in terms of the total mass density by the equation of state of an ideal gas. With these profiles, PakalMPI computes the abundances and electron densities of the atmospheric model similar to \cite{Vernazza_et_al_1973}, but in our case the Saha equation is solved for the base state and for high energy levels of each ion. Details of calculations can be found in \cite{De_la_Luz_et_al_2011}. 
 
\begin{figure*}
\centering
\includegraphics[width=1.0\textwidth]{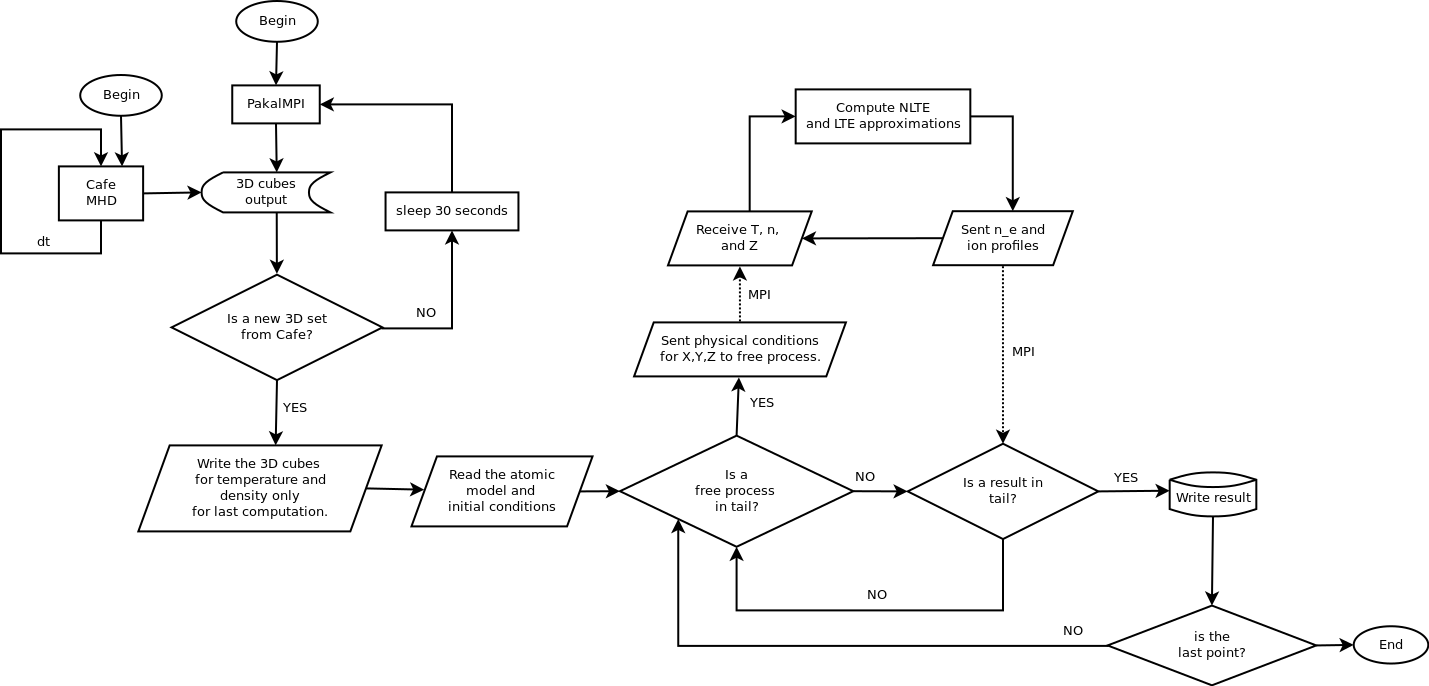}
\caption{Diagram of CAFE-PakalMPI model.}
\label{general_diagram}
\end{figure*}

 The general diagram of CAFE-PakalMPI model is shown in Figure \ref{general_diagram}, here we describe two processes working together. The first step computes the physical conditions of the system using the MHD approximation, the results are written in a file in form of a 3D cube. The second step check this file for changes, if there is a new set of data then we split the file into subset of 3D cubes for temperature ($T$) and density ($\rho$), then read into RAM memory the correspond pair of 3D cubes ($T$ and $\rho$) to be computed into PakalMPI. Each spatial point in the pair of 3D cubes is transformed in a request in MPI to be solved in the pool of available process. Each process have a copy of PakalMPI NLTE solver that only receives a single $T$ and a single $\rho$. Each PakalMPI process computes HI, HII, HeI, HeII, HeIII, H-, $n_e$, and the other species for the particular physical condition. Each process send back the result for the particular point and return to the state of wait for another job. The master process recovery the results on the fly and store the results in the hard disk.

\section{Results}
\label{results}

\begin{figure*}
\centering
\includegraphics[width=8.0cm,height=5.0cm]{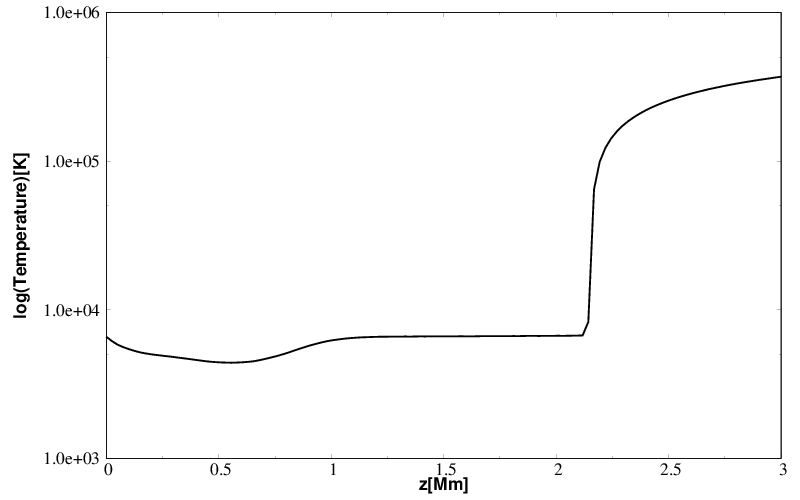}
\includegraphics[width=8.0cm,height=5.0cm]{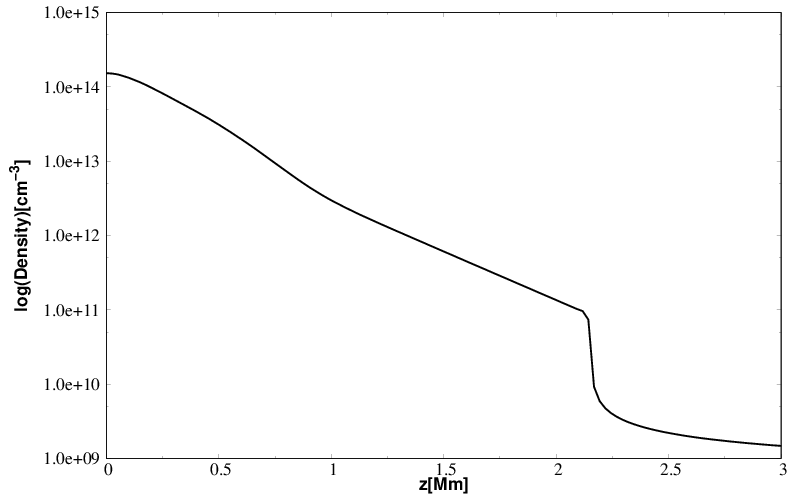}\\
\includegraphics[width=8.0cm,height=5.0cm]{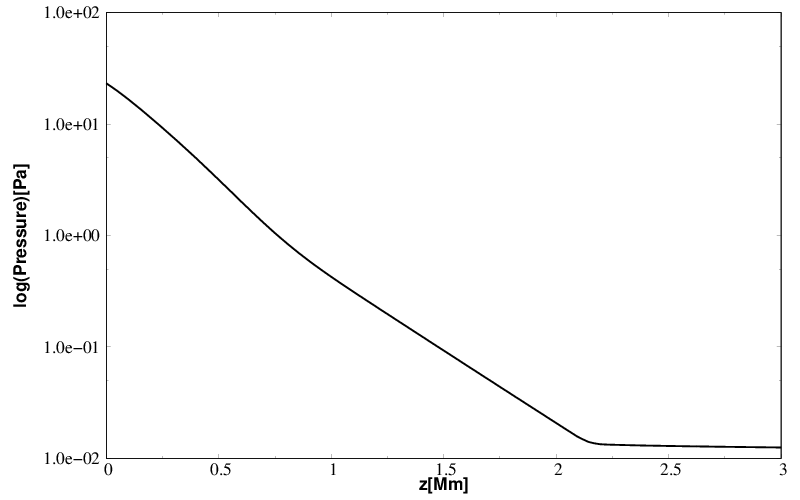}
\includegraphics[width=8.0cm,height=5.0cm]{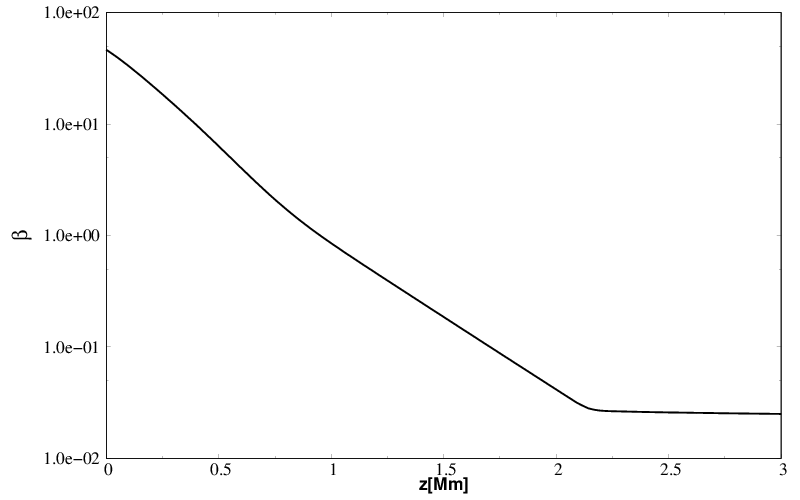}
\caption{(Top) Logarithm of temperature (K) and total mass density $\rho$(cm$^{-3}$) as a function of height $z$ at initial time for the C7 equilibrium model. (Bottom) Logarithm of gas pressure (Pa) and plasma beta as a function of height $z$ at initial time. Notice the steep jump at transition region $z\approx$2.1 Mm.}
\label{C7_hydrostatic}
\end{figure*}

We apply CAFE-PakalMPI to study the stability of a solar atmosphere in magnetohydrostatic equilibrium. In this case, we consider that the temperature field is assumed to obey the semi-empirical C7 model of the chromosphere-transition region, e.g., \citep{Avrett&Loeser2008}. The model includes the photosphere as presented in \citep{Fontela_et_al_1990}. The magnetic field configuration is described by a straight field $B_z =$ 11.21 G in the vertical direction. The profiles at initial time of $T(z)$, $\rho(z)$, $p(z)$ and plasma beta are shown in Figure \ref{C7_hydrostatic}, these profiles represent a good approximation of the conditions of the low solar atmosphere \citep{Gary_2001}. 

In this test, we define a specific domain covering part of the photosphere, chromosphere, transition region and corona. For this, we define the numerical domain to be $x\in[0,3]$, $y\in[0,3]$, $z\in[0,3]$ Mm, covered with 120$\times$120$\times$120 grid cells, i.e., the effective resolution is 25 km in each direction. We solve the MHD equations with CAFE using the HLLE flux formula combined with the MINMOD reconstructor and a constant Courant factor CFL = 0.2. We use fixed in time boundary conditions at $z=0$ Mm, which means that the values of the conservative variables are set to their initial condition values at this boundary, in the other faces we use out-flux boundary conditions \citep{Gonzalez_Aviles_et_al_2018a}.

We set the atmospheric model and the magnetic field configuration described above, then we use CAFE to evolve the plasma according to the ideal MHD equations under the effect of a solar constant gravitational field. Since the system is initially in magnetohydrostatic equilibrium, it should remain static forever, however due to numerical diffusion of the numerical scheme adopted, the configuration will slowly dissipate. In this case, for simplicity we chose a simulation time between the time formation of a Type II spicule $t\approx 100$ s \citep{De_Pontieu_et_al_2007, Pereira_et_al_2012}, and the five minute (300 s) oscillations in the solar atmosphere \citep{Khomenko_et_al_2008,Zaqarashvili_et_al_2011}. 

Once we obtained the plasma mass density and temperature with CAFE, we use CAFE-PakalMPI to obtain the 3D profiles of the ionized Hydrogen density HI(cm$^{-3}$), singly ionized Hydrogen density HII(cm$^{-3}$), singly ionized Helium density HeII(cm$^{-3}$), doubly ionized Helium density HeIII(cm$^{-3}$), negative ionized Hydrogen density H-(cm$^{-3}$) and the electronic density $n_e$(cm$^{-3}$).  

\begin{figure*}
\centering
\includegraphics[width=5.0cm,height=3.0cm]{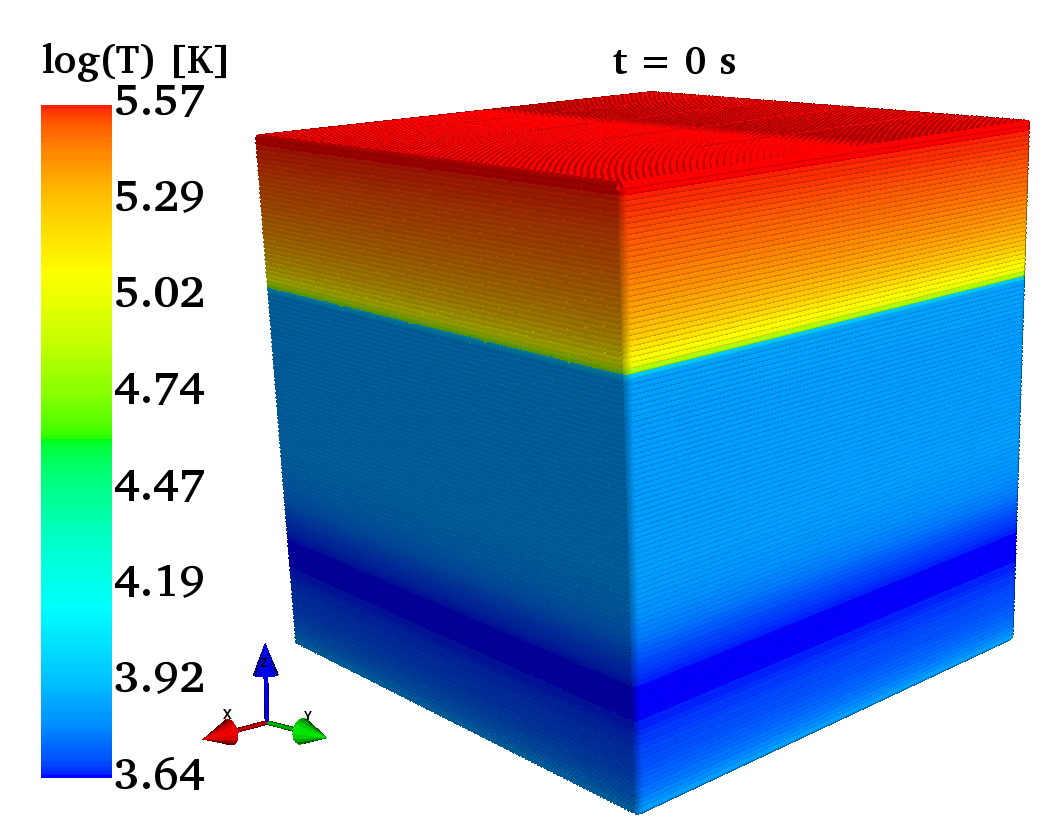}
\includegraphics[width=5.0cm,height=3.0cm]{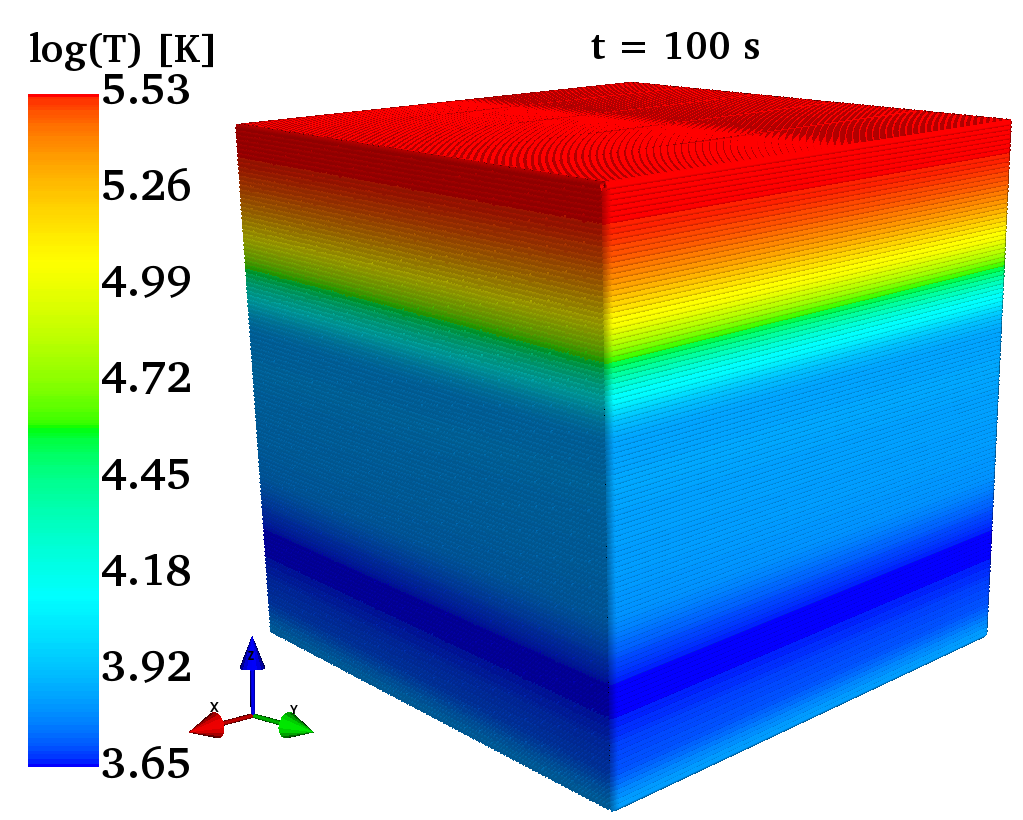}
\includegraphics[width=5.0cm,height=3.0cm]{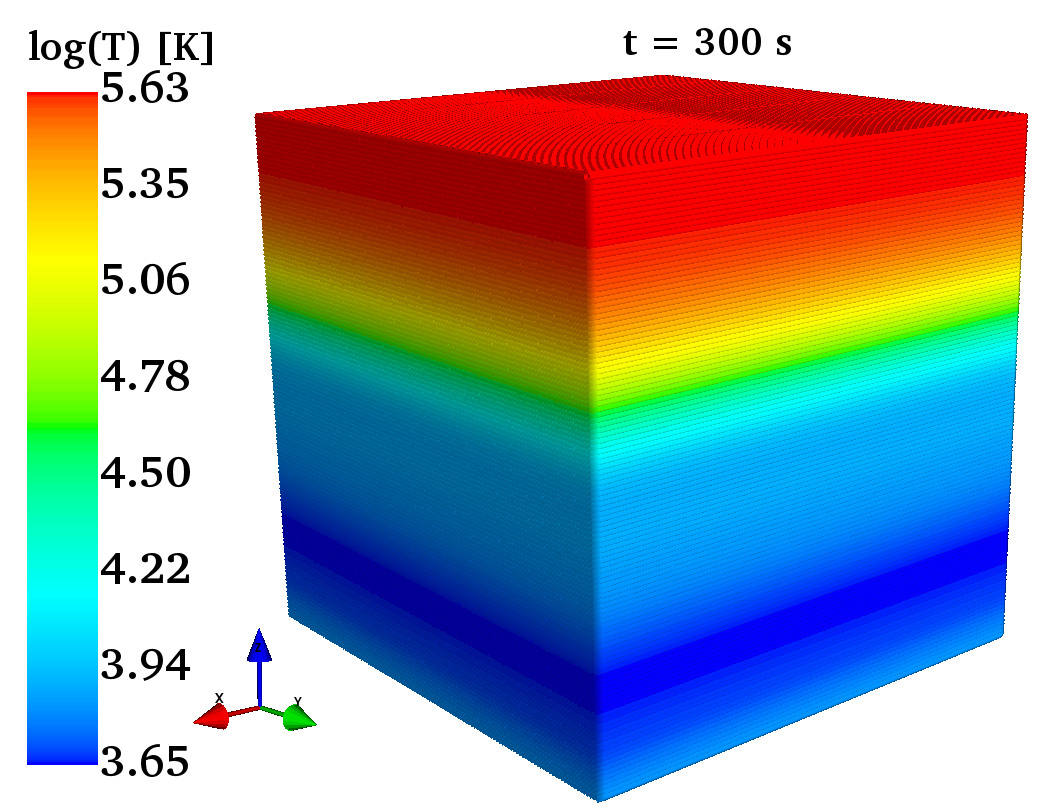}\\
\includegraphics[width=5.0cm,height=3.0cm]{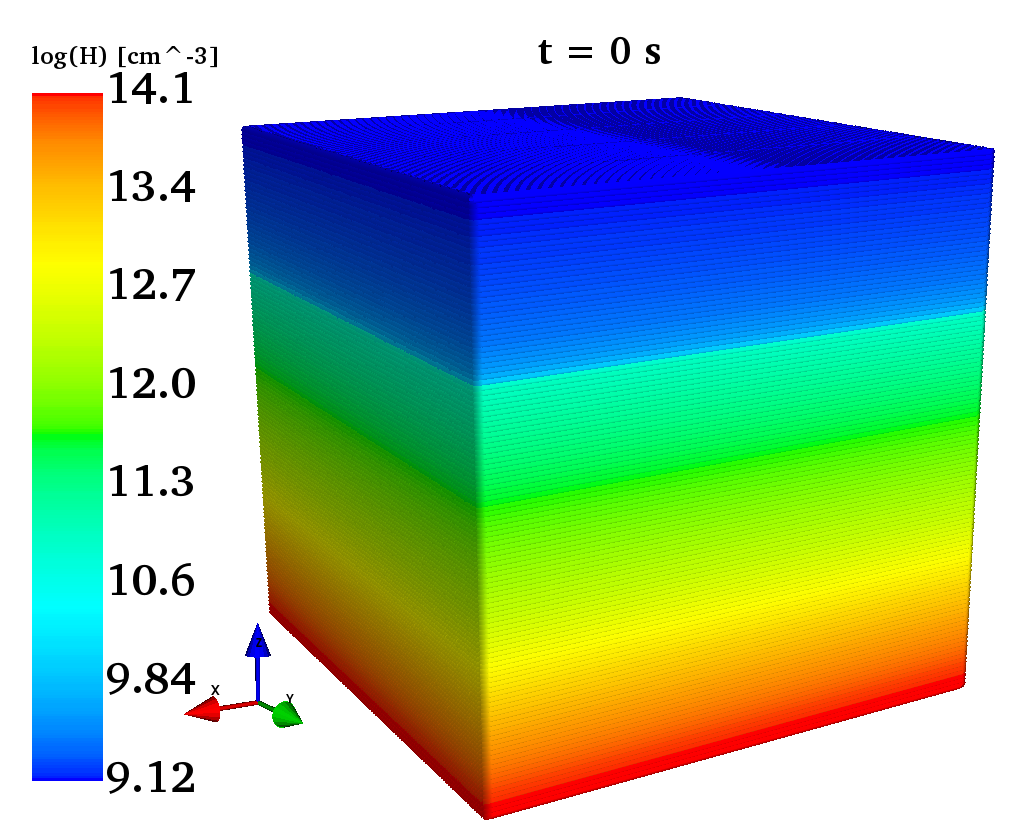}
\includegraphics[width=5.0cm,height=3.0cm]{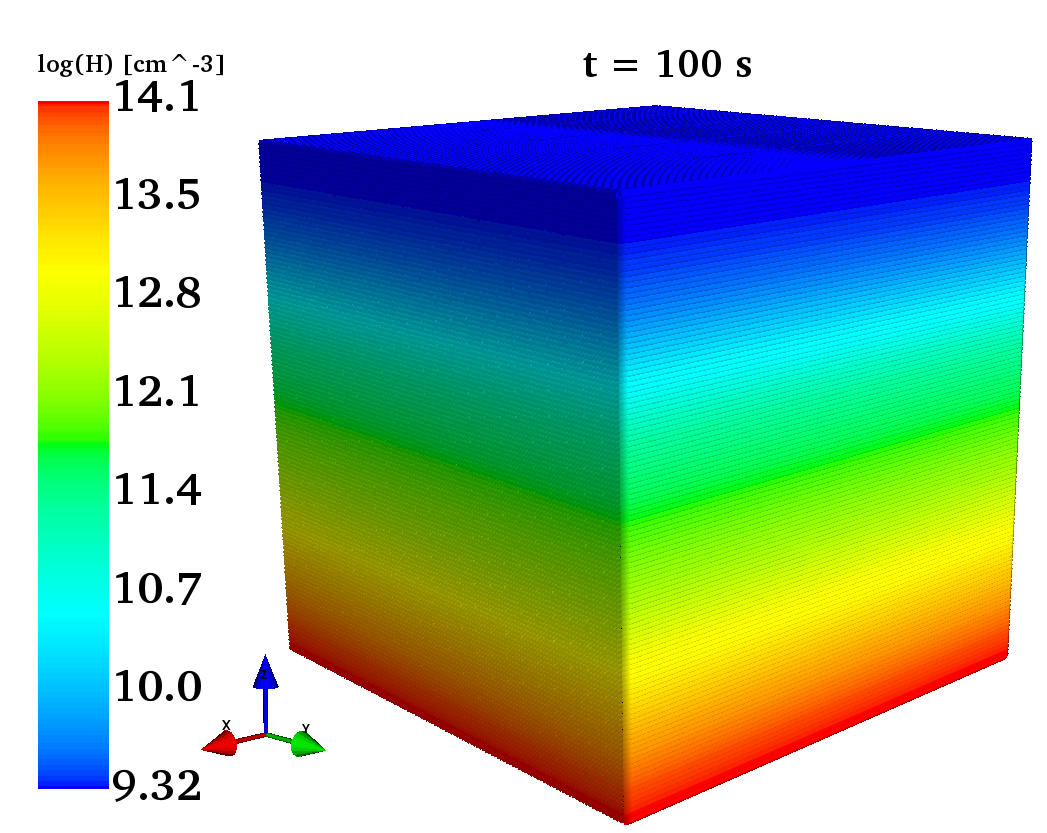}
\includegraphics[width=5.0cm,height=3.0cm]{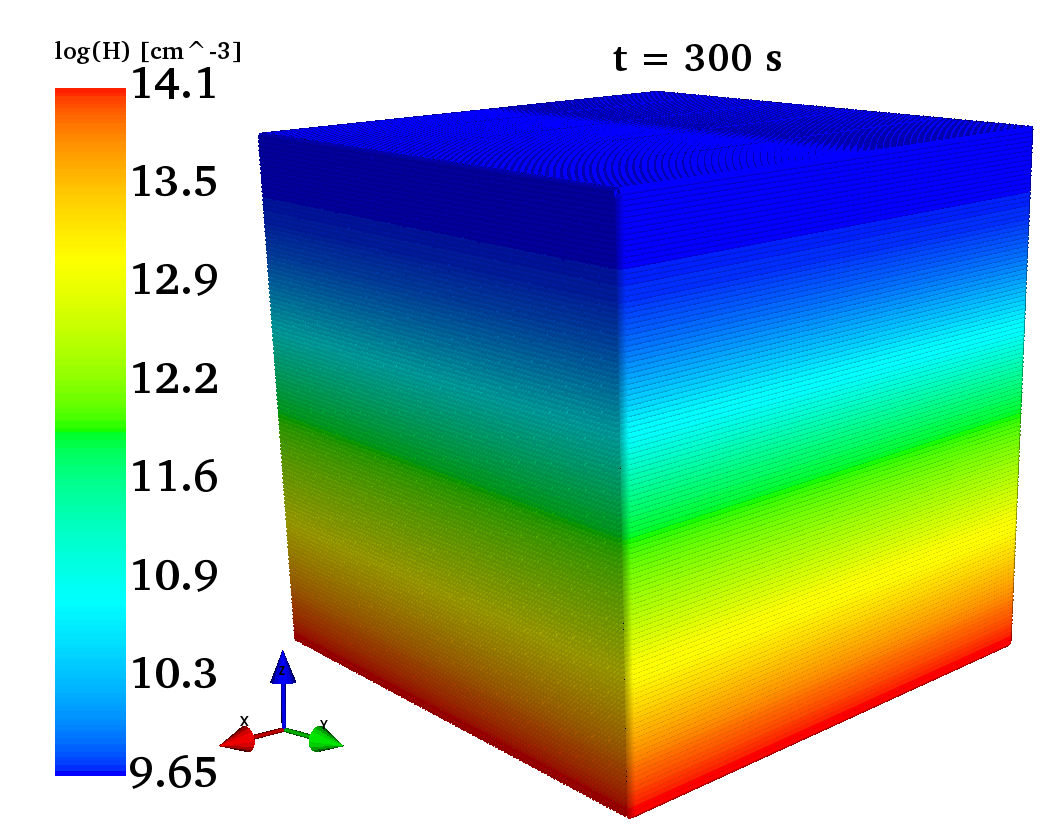}
\caption{(Top) 3D representation of the logarithm of temperature in Kelvin at times 0, 100 and 300 s. (Bottom) 3D representation of the logarithm of total Hydrogen density H(cm$^{-3}$ at times 0, 100 and 300 s.}
\label{initial_conditions_3D_temp_H}
\end{figure*}

\begin{figure*}
\centering
\includegraphics[width=5.0cm,height=3.0cm]{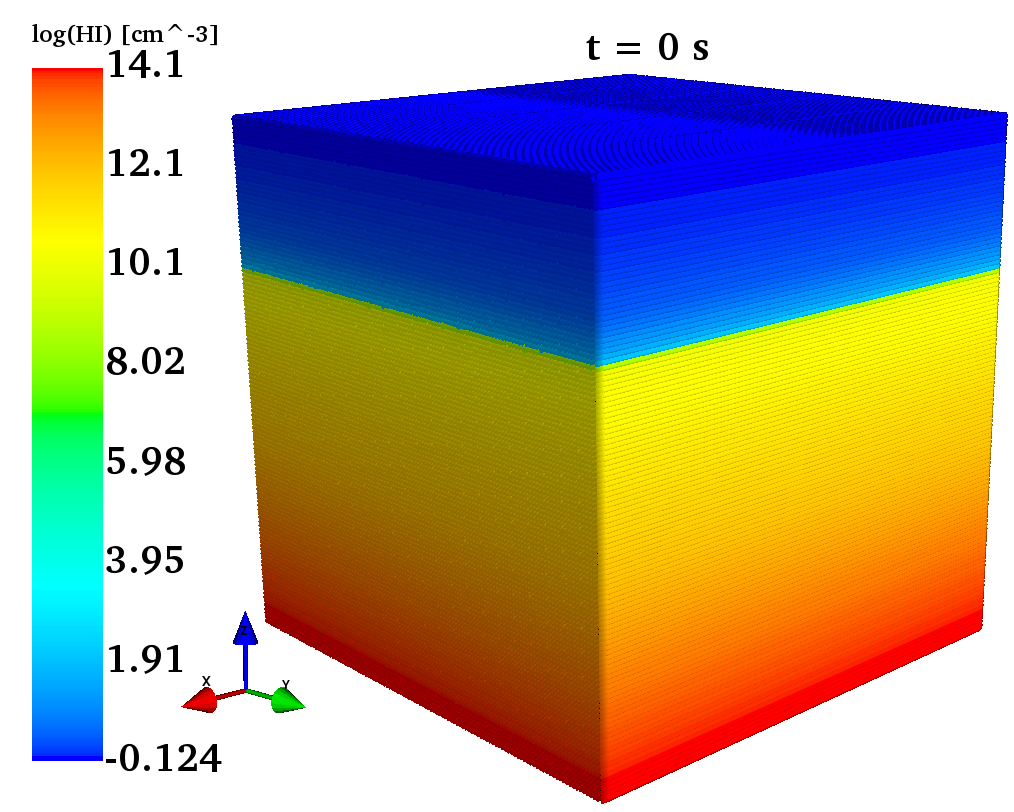}
\includegraphics[width=5.0cm,height=3.0cm]{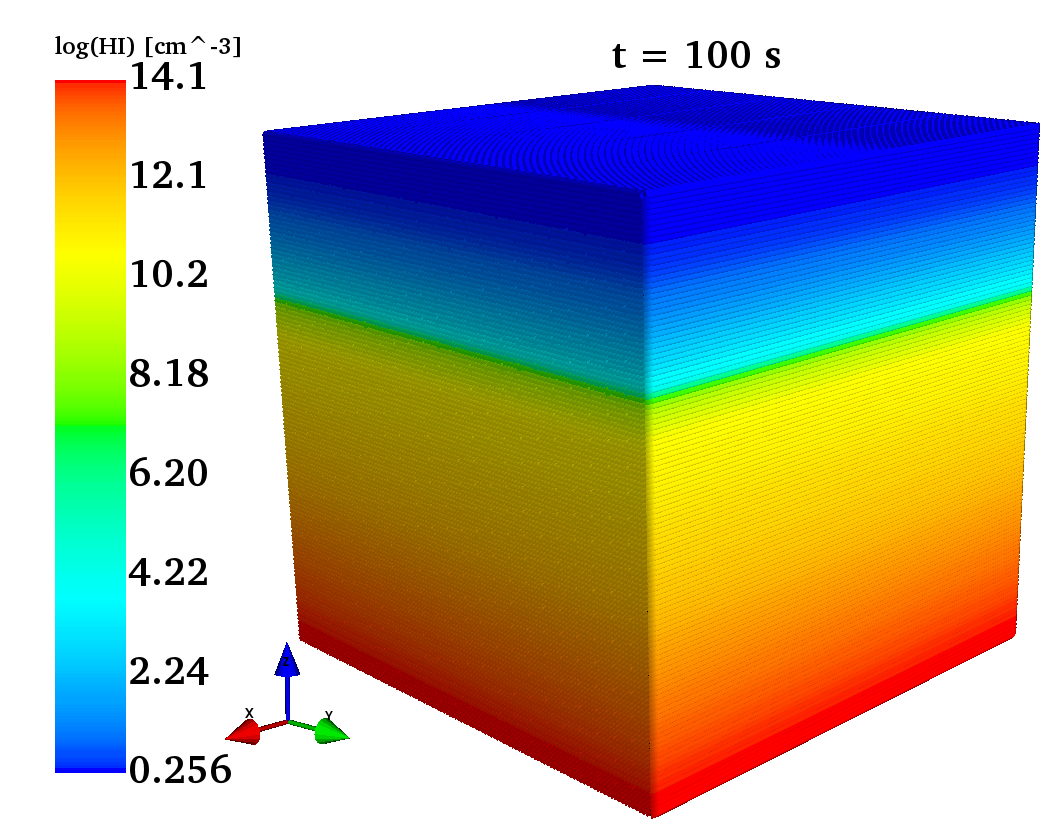}
\includegraphics[width=5.0cm,height=3.0cm]{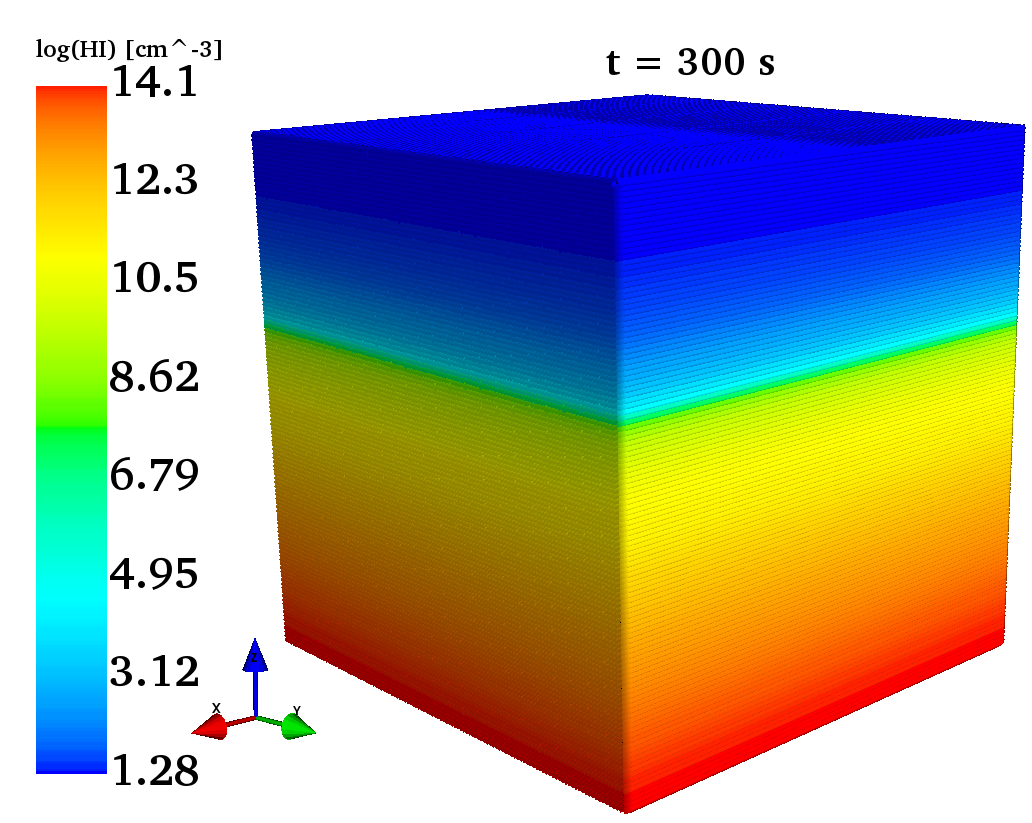}\\
\includegraphics[width=5.0cm,height=3.0cm]{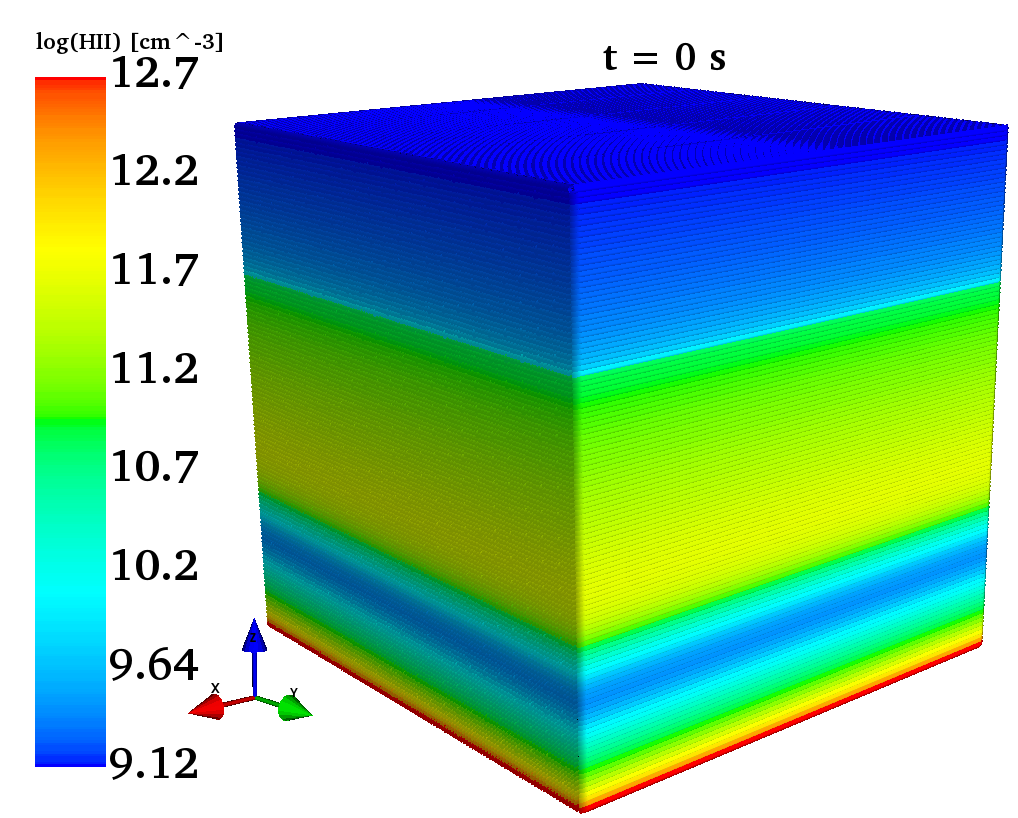}
\includegraphics[width=5.0cm,height=3.0cm]{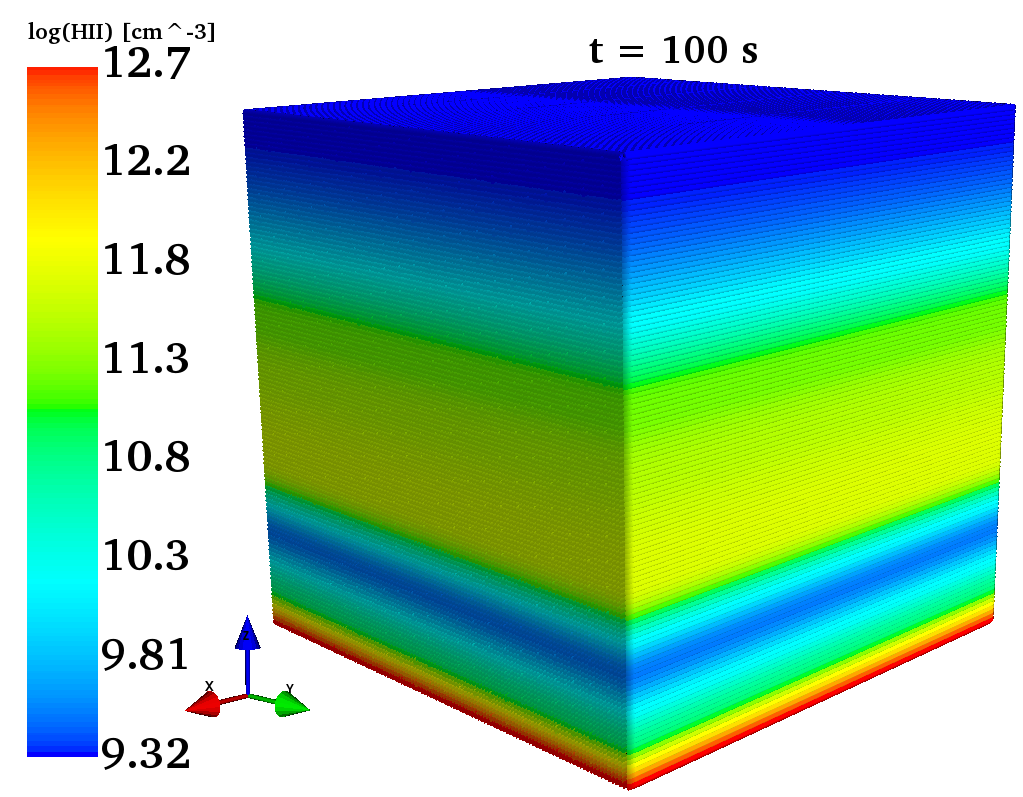}
\includegraphics[width=5.0cm,height=3.0cm]{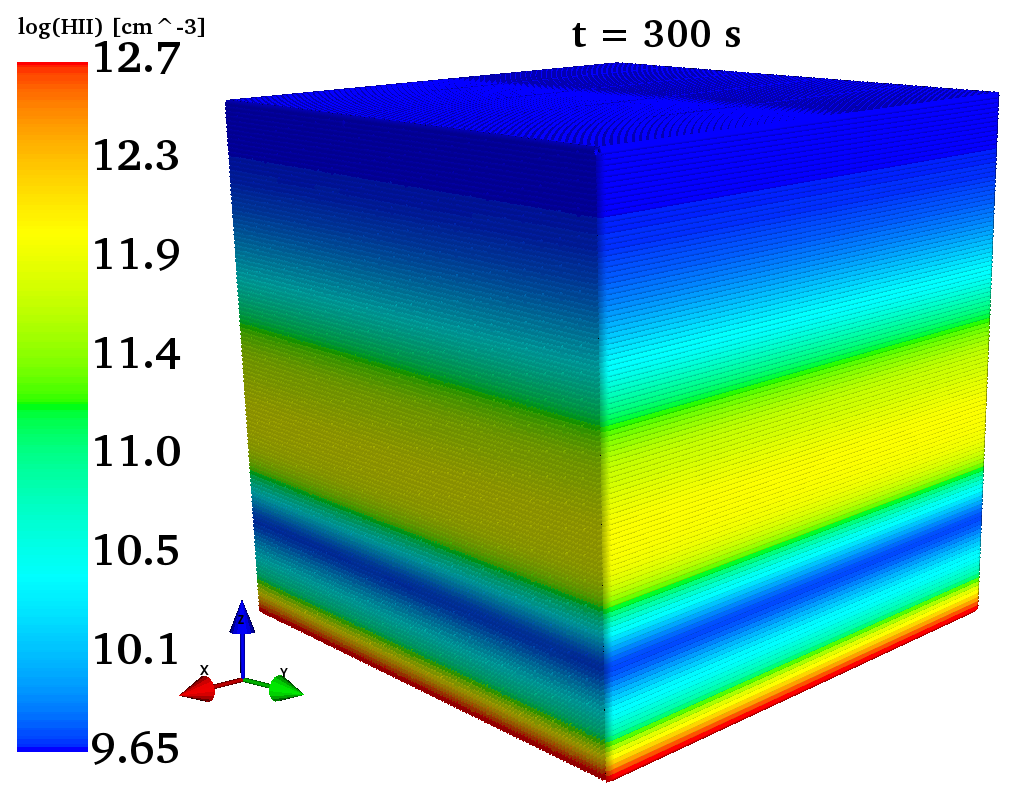}
\caption{(Top) 3D representation of the logarithm of ionized Hydrogen density HI in cm$^{-3}$ at times 0, 100 and 300 s. (Bottom) 3D representation of the logarithm of singly ionized Hydrogen density HII in cm$^{-3}$ at times 0, 100 and 300 s.}
\label{3D_evolution_HI_HII}
\end{figure*}

\begin{figure*}
\centering
\includegraphics[width=5.0cm,height=3.0cm]{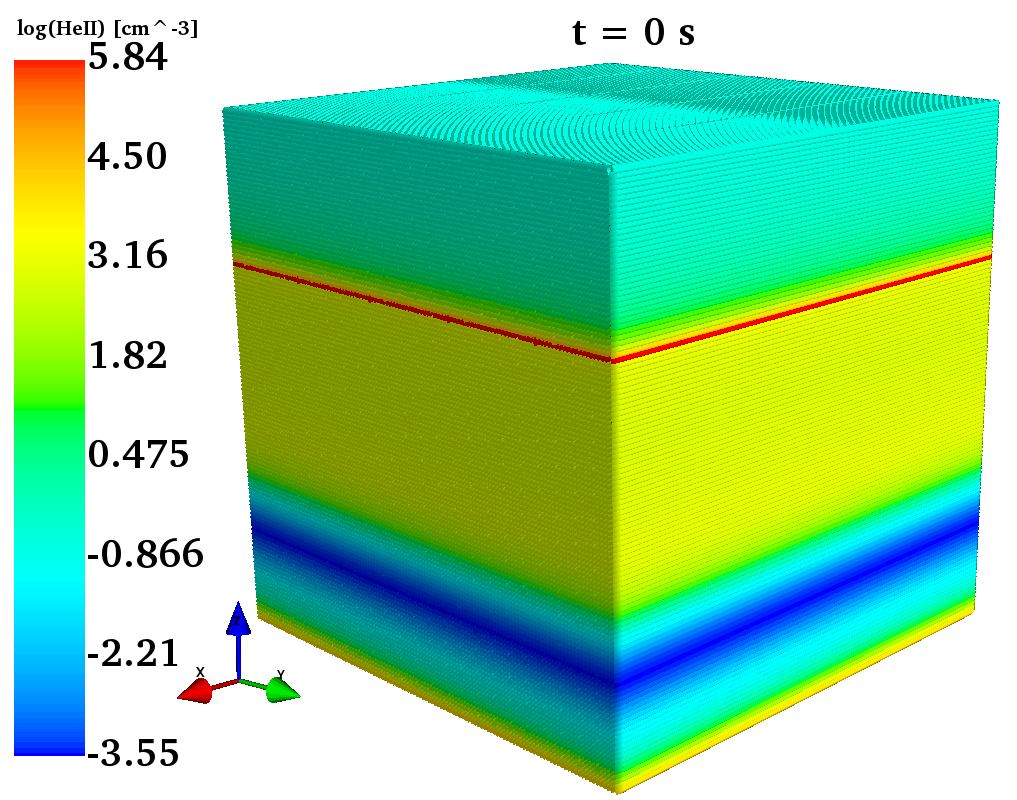}
\includegraphics[width=5.0cm,height=3.0cm]{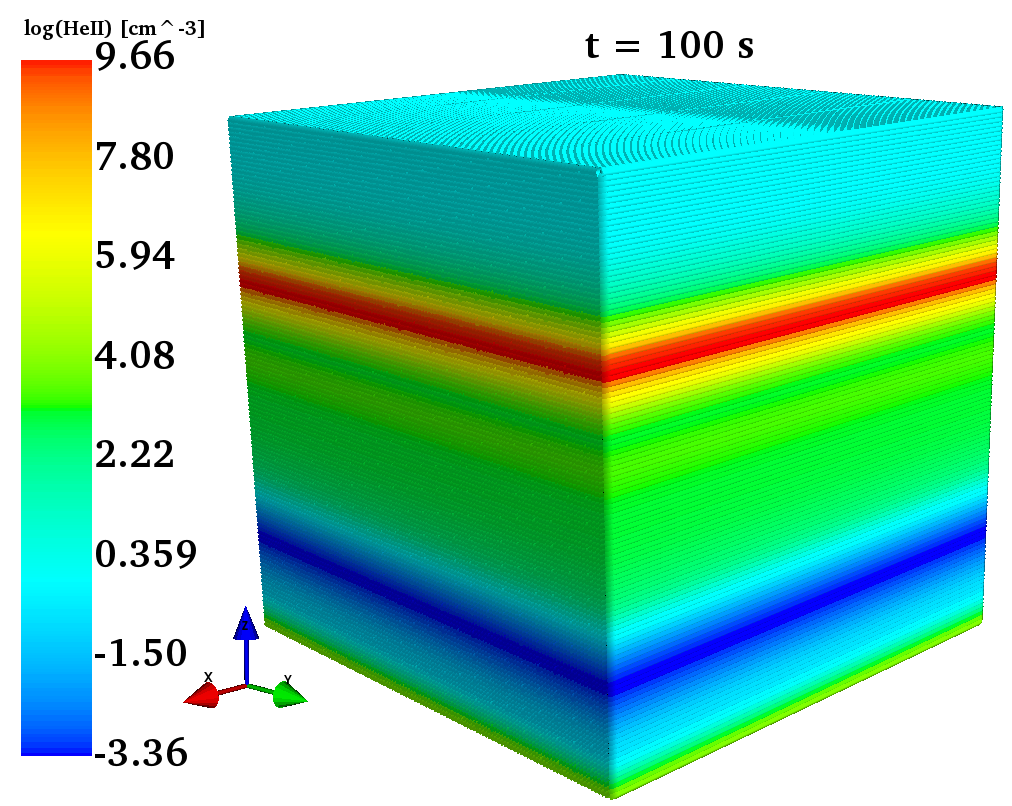}
\includegraphics[width=5.0cm,height=3.0cm]{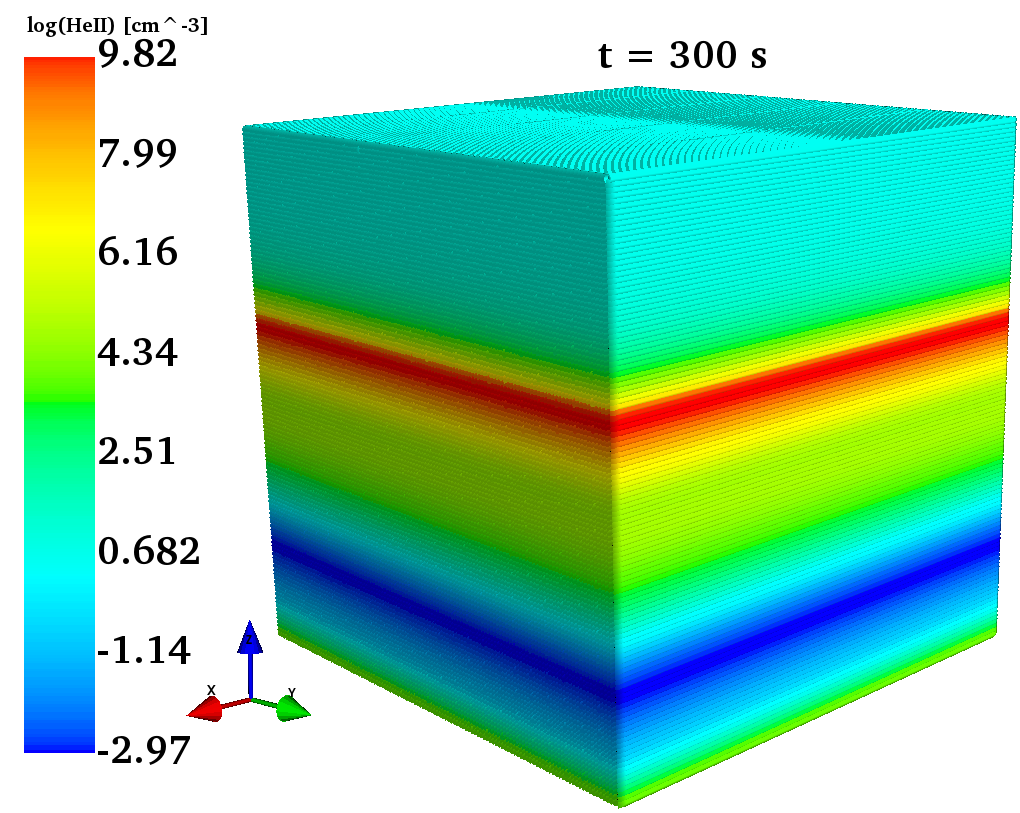}\\
\includegraphics[width=5.0cm,height=3.0cm]{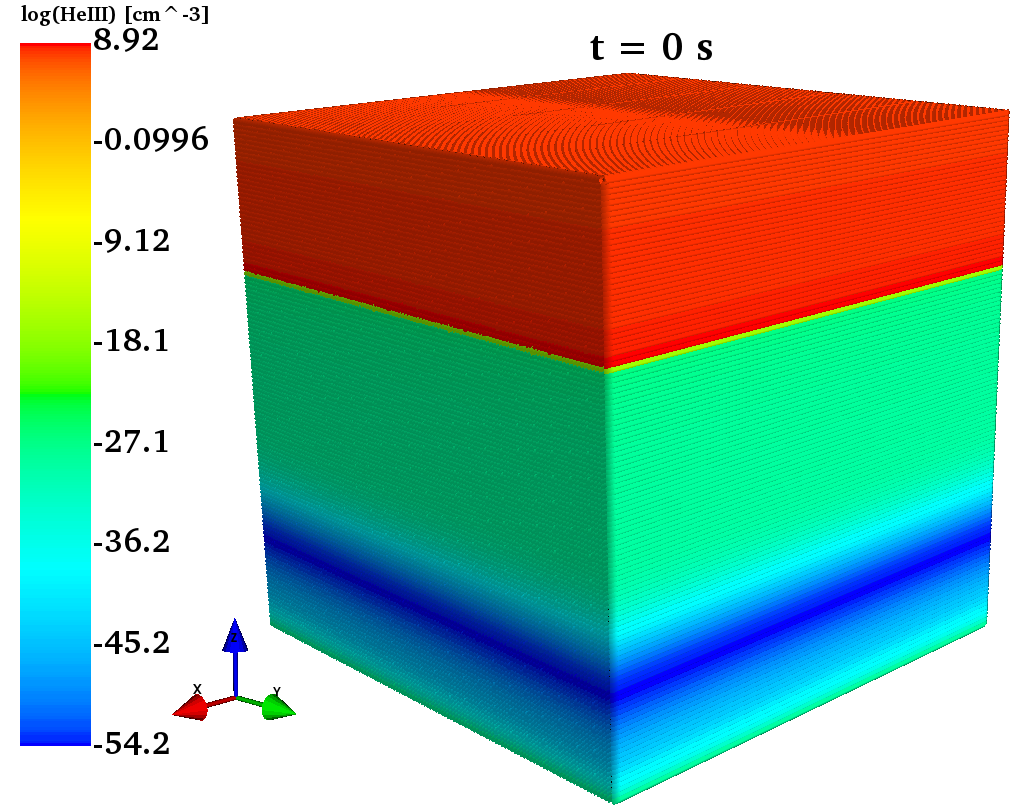}
\includegraphics[width=5.0cm,height=3.0cm]{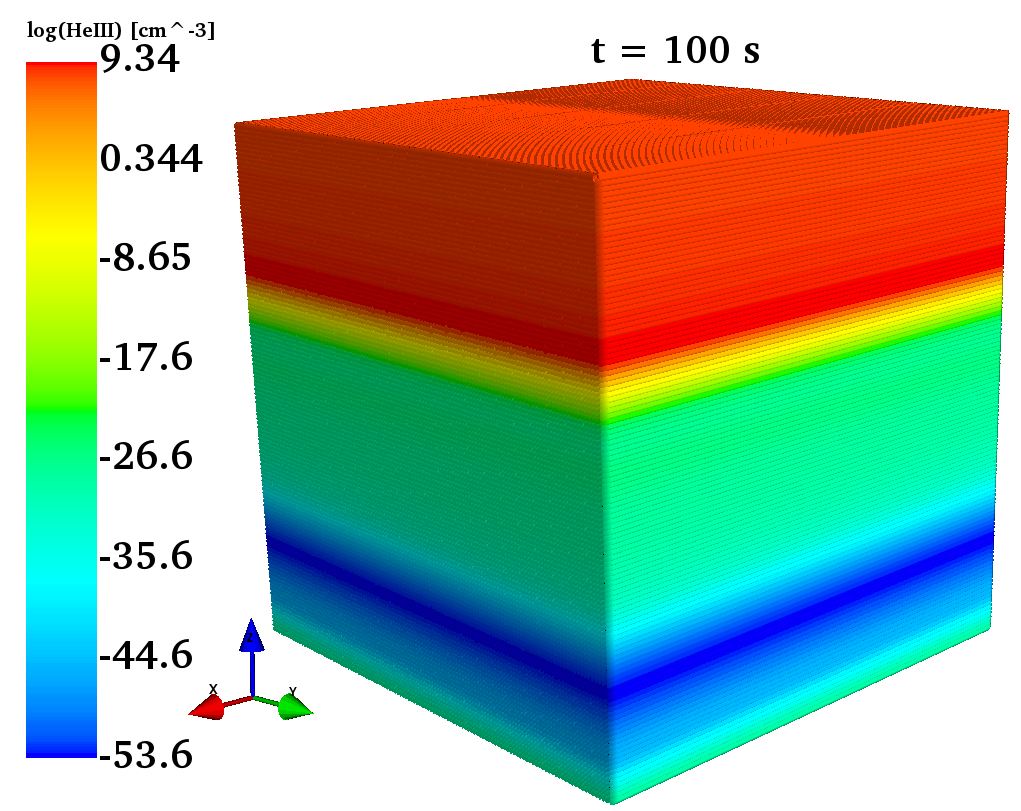}
\includegraphics[width=5.0cm,height=3.0cm]{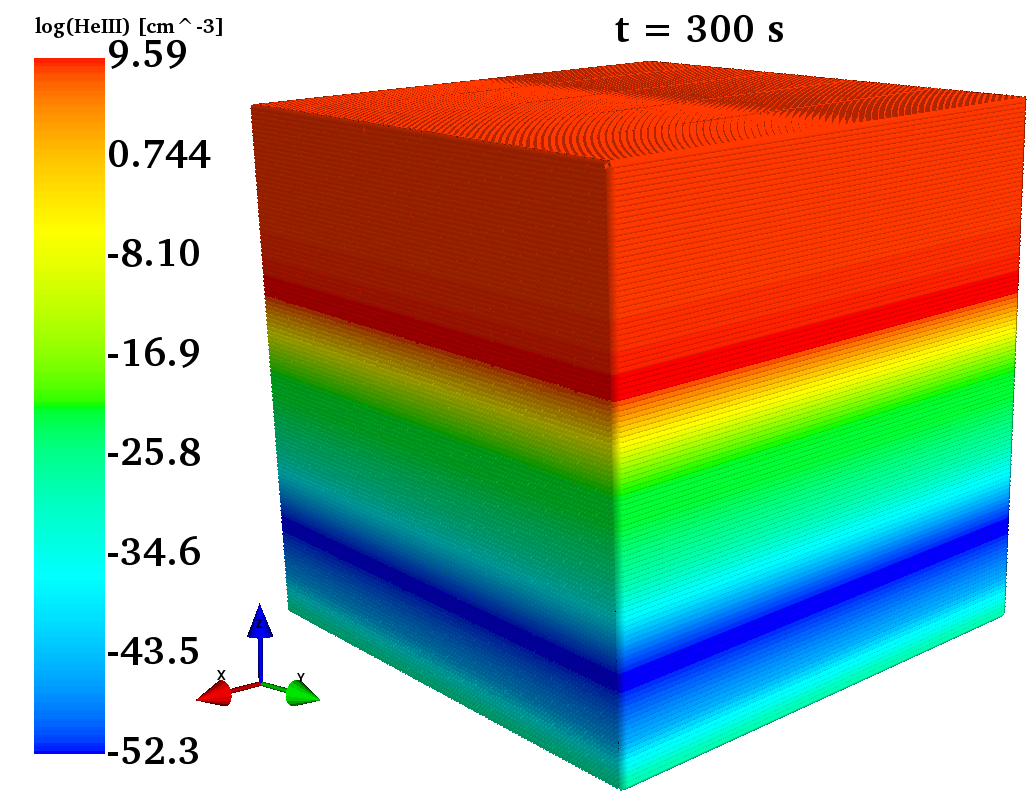}
\caption{(Top) 3D representation of the logarithm of singly ionized Helium density HeII (cm$^{-3}$) at times 0, 100 and 300 s. (Bottom) 3D representation of the logarithm of doubly ionized Helium density HeIII (cm$^{-3}$) at times 0, 100 and 300 s.}
\label{3D_evolution_HeII_HeIII}
\end{figure*}

\begin{figure*}
\centering
\includegraphics[width=5.0cm,height=3.0cm]{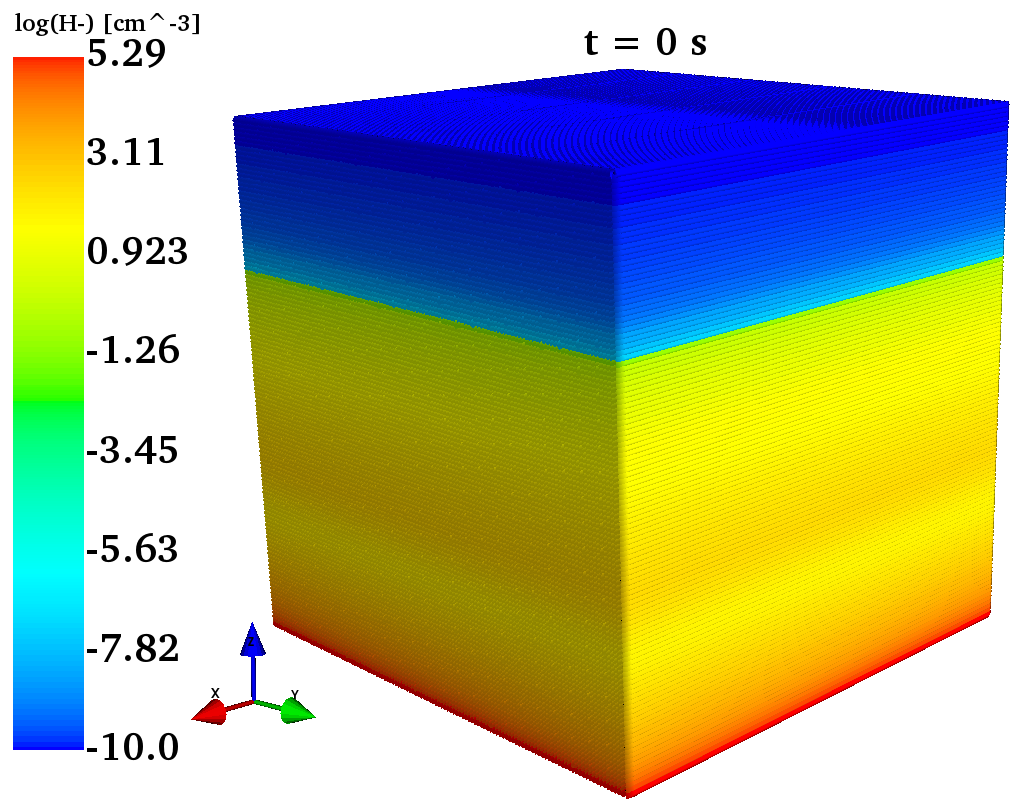}
\includegraphics[width=5.0cm,height=3.0cm]{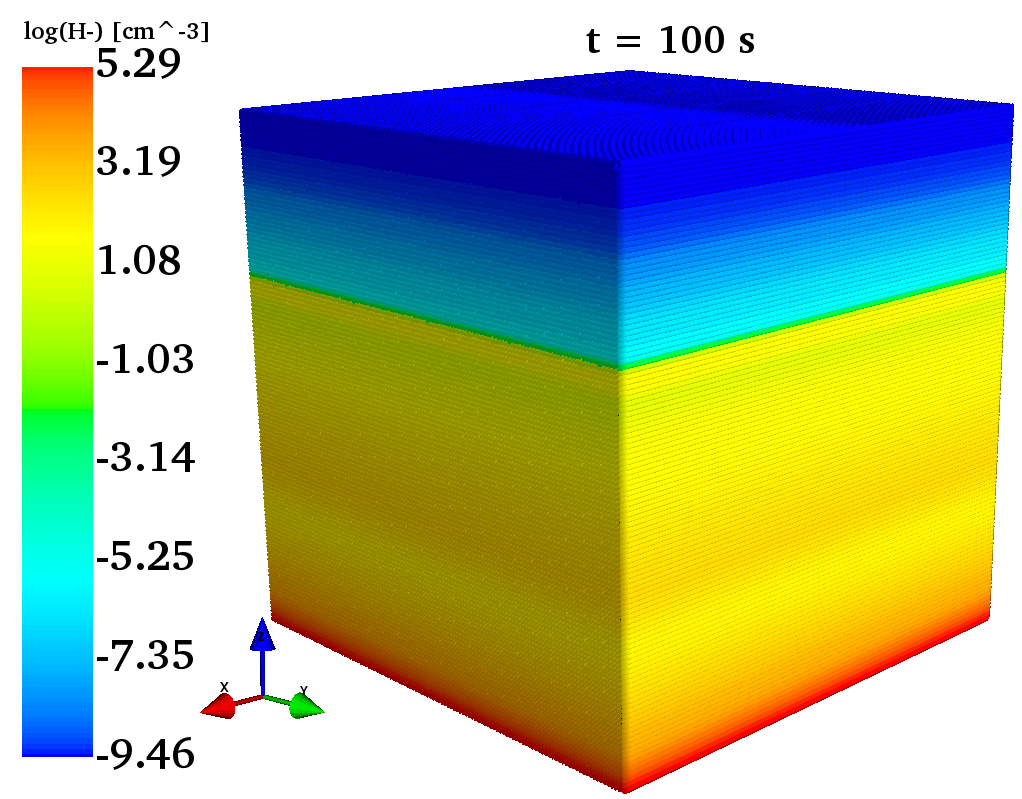}
\includegraphics[width=5.0cm,height=3.0cm]{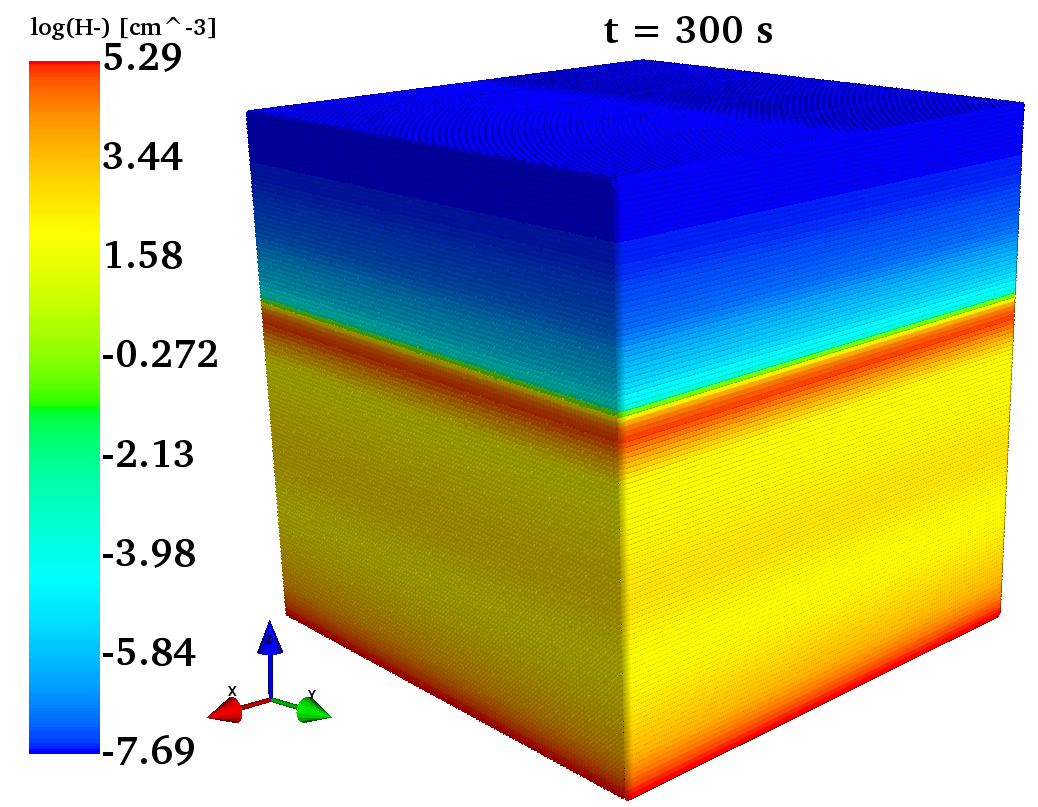}\\
\includegraphics[width=5.0cm,height=3.0cm]{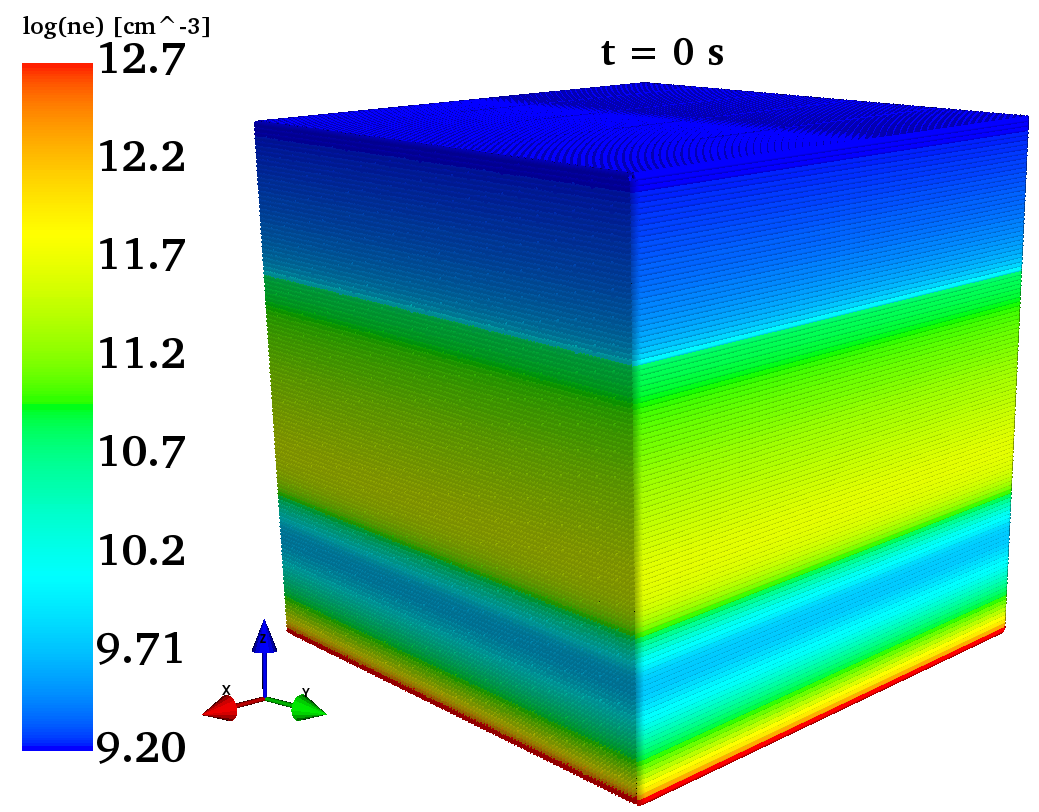}
\includegraphics[width=5.0cm,height=3.0cm]{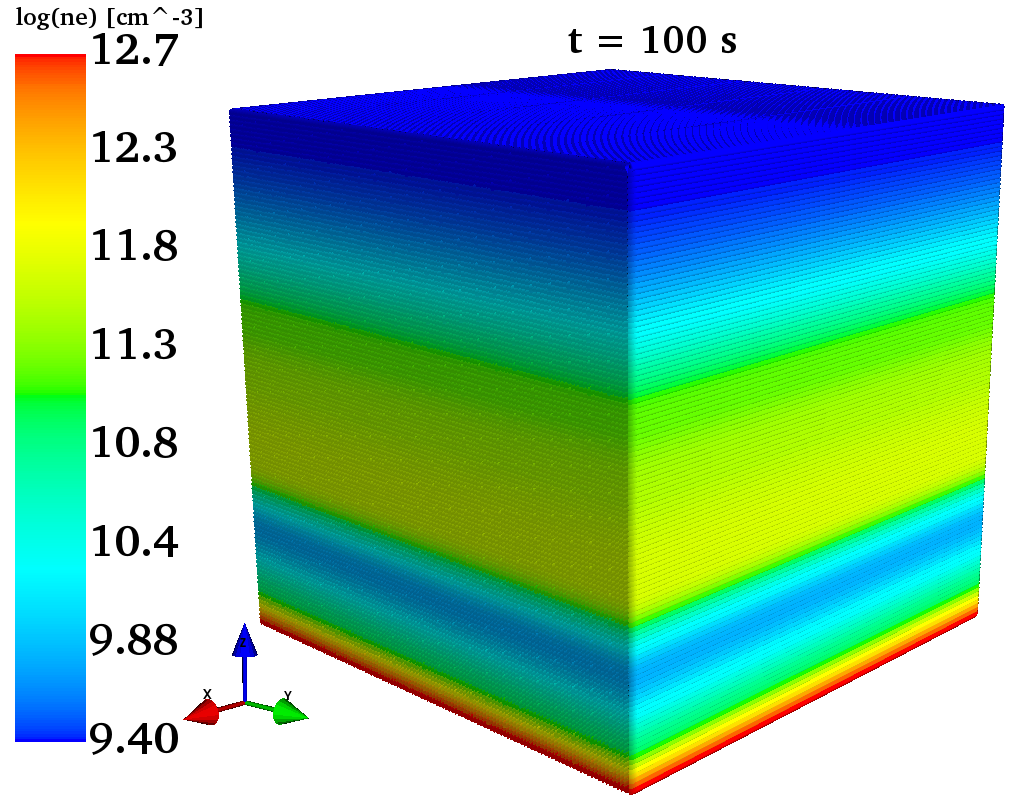}
\includegraphics[width=5.0cm,height=3.0cm]{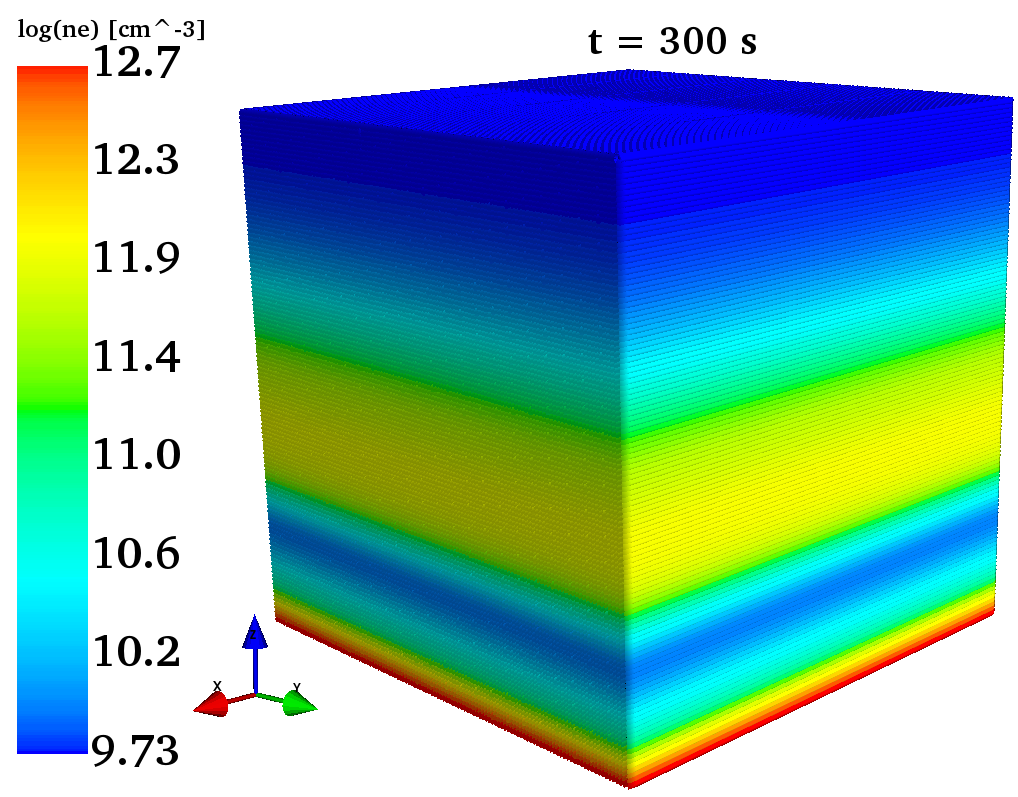}
\caption{(Top) 3D representation of the logarithm of negative ionized Hydrogen density H- (cm$^{-3}$) at times 0, 100 and 300 s. (Bottom) 3D representation of the logarithm of electronic density $n_e$ (cm$^{-3}$) at times 0, 100 and 300 s.}
\label{3D_evolution_H-_ne}
\end{figure*}

For instance, in Figure \ref{initial_conditions_3D_temp_H}, we show snapshots of the logarithm of temperature (K) and total Hydrogen density H (cm$^{-3}$) at three times $t=0$, $t=100$ s and $t=300$ s. In both cases we see that the profiles are stable at the bottom and top of the atmosphere, however near to the transition region $z\approx$2.1 Mm, we can see diffusion of plasma. In Figure \ref{3D_evolution_HI_HII}, we show snapshots of the logarithm of the ionized Hydrogen density HI(cm$^{-3}$) and singly ionized Hydrogen density HII(cm$^{-3}$) at times $t=0$, $t=100$ and $t=300$ s. In this case, we can see that both profiles remain stable, however they show diffusion near to the transition region. In addition, in Figure \ref{3D_evolution_HeII_HeIII} we show snapshots of the singly ionized Helium density HeII(cm$^{-3}$) and of the doubly ionized Helium density HeIII(cm$^{-3}$) at times $t=0$, $t=100$ and $t=300$ s. In this case, we see diffusion at about $z\approx$ 0.5 Mm and near to the transition region, despite these regions the system remains stable at bottom and top of the atmosphere. Finally, in Figure \ref{3D_evolution_H-_ne}, we show snapshots of the logarithm of negative ionized Hydrogen density H-(cm$^{-3}$) and of the electronic density $n_e$(cm$^{-3}$) at the same three times mentioned above, in this case the diffusion is also localize in the transition region. 

\begin{figure*}
\centering
\includegraphics[width=8.0cm,height=5cm]{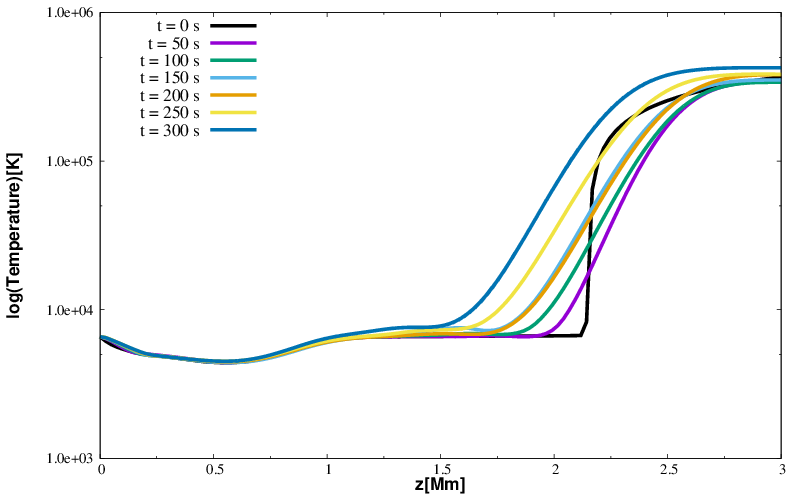}
\includegraphics[width=8.0cm,height=5.0cm]{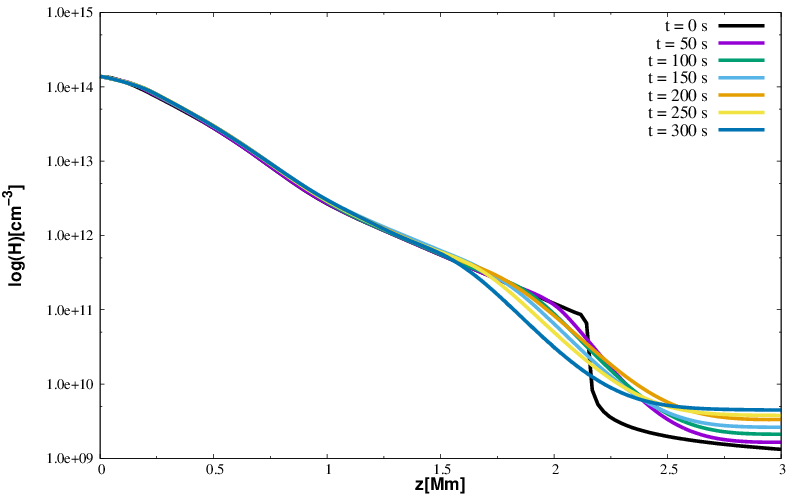}
\includegraphics[width=8.0cm,height=5.0cm]{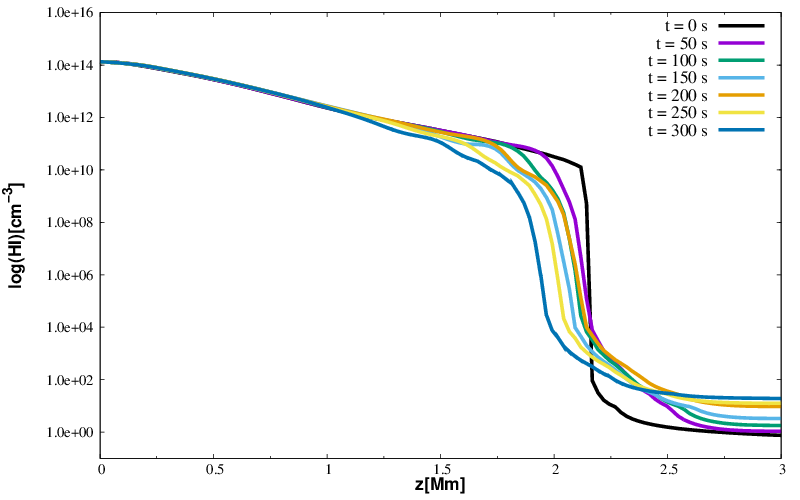}
\includegraphics[width=8.0cm,height=5cm]{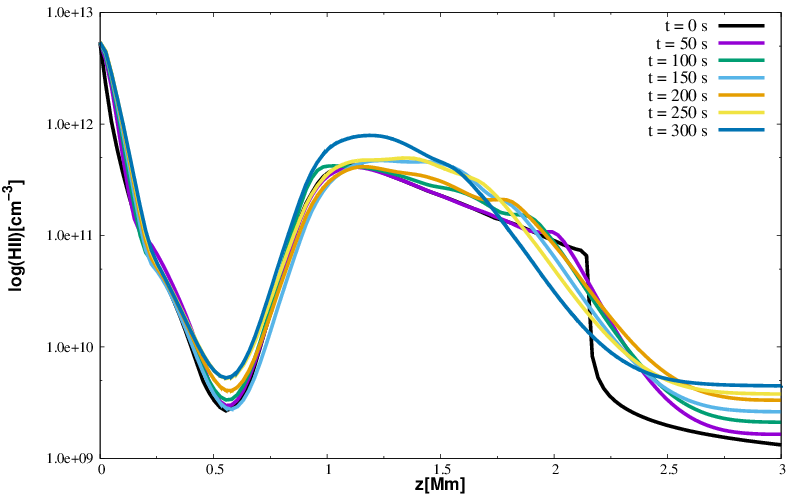}
\includegraphics[width=8.0cm,height=5.0cm]{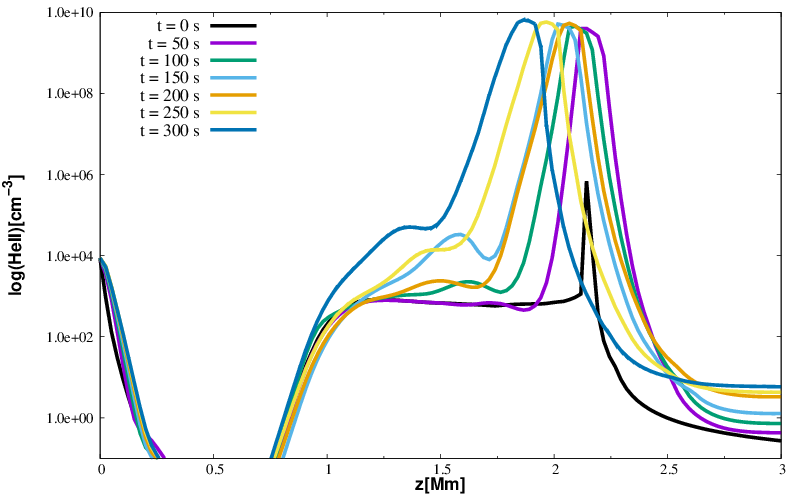}
\includegraphics[width=8.0cm,height=5cm]{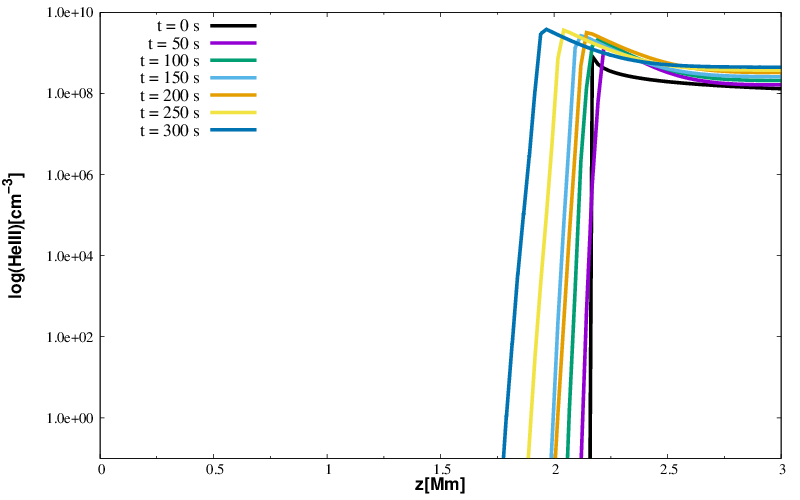}
\includegraphics[width=8.0cm,height=5.0cm]{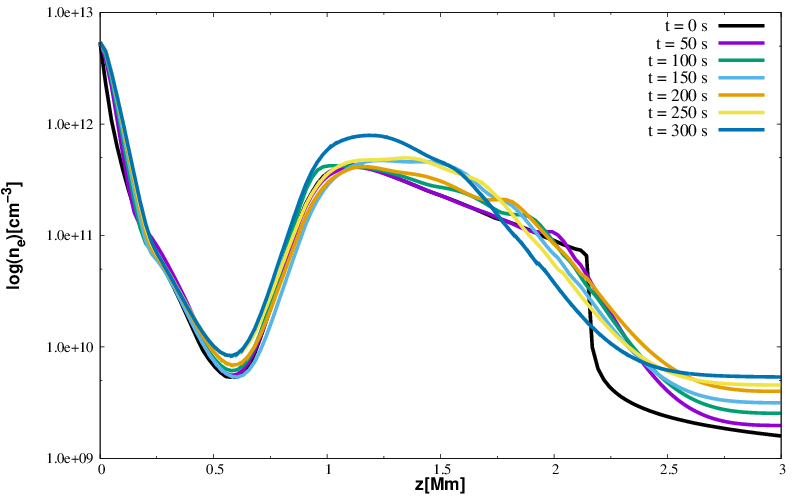}
\includegraphics[width=8.0cm,height=5cm]{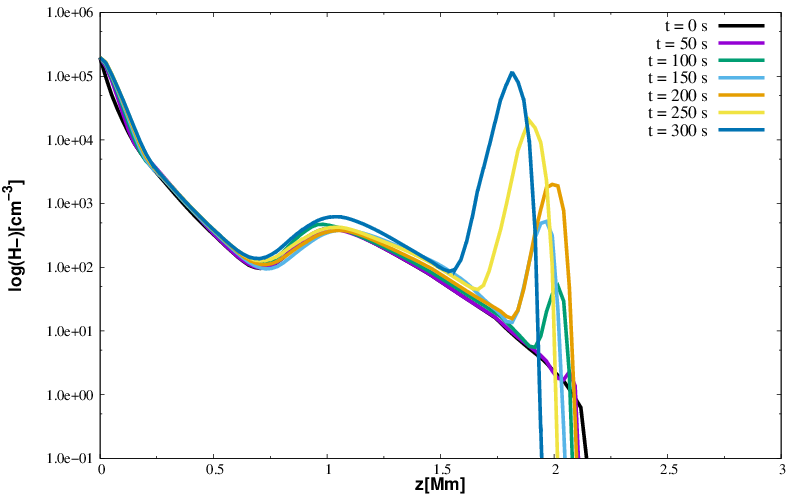}
\caption{(Top-panel) Snapshots of temperature (K) and total Hydrogen density H(cm$^{-3}$) as function of $z$ for various times. (Second panel) Snapshots of ionized Hydrogen HI (cm$^{-3}$) and singly ionized Hydrogen density HII (cm$^{-3}$) as function of $z$ for various times. (Third panel) Snapshots of singly ionized Helium density HeII (cm$^{-3}$) and doubly ionized Helium density HeIII (cm$^{-3}$) as function of $z$ for various times. (Bottom-panel) Snapshots of electronic density $n_{e}$ (cm$^{-3}$) as function of $z$ and negative ionized Hydrogen density H- (cm$^{-3}$) as function of $z$ for various times. Note the waves formed in the different profiles.}
\label{evolution_of_ionization_states}
\end{figure*}

In the 3D representation of the density profiles we can not see clearly how diffusion occurs near to the transition region, for this reason we obtain the profiles as a function of the height $z$ for various times. In the first panel of Figure \ref{evolution_of_ionization_states}, we show the profile of the temperature ($T$) and the total Hydrogen density (H), in this case we can see that temperature remains stable until $z\approx 1$ Mm, just before of the Temperature Minimum region (TMR) (observed between 0 and 1 Mm). After the TMR, the set of temperature shows an slight increase around 1.5 Mm and then a marked diffusion at the transition region that extends between 1.5 and 2.75 Mm. 
In the case of H, we can see a stable behavior until $z\approx$1.5 Mm, after this altitude over the photosphere the density decrease faster each time step, however this behavior is reverted just before at 2.5Mm. In principle, the stable behavior near to the photosphere could be related to the fixed boundary condition at $z=0$ Mm and the diffusion by the steep gradient presented at the transition region. 

In the second panel of Figure \ref{evolution_of_ionization_states}, we show the ionized Hydrogen density HI and the singly ionized Hydrogen density HII. In the case of step zero, HI drops dramatically at 2.25 Mm (HI breaking point), this altitude corresponds to the temperature increase in the transition region. The altitude of this breaking point drops with the time steps due to the changes in temperature of our model. The result is that the density of HII increases at lower altitudes for each time step. We also note that both plots show a wave that seem start at the HI breaking point and then travel in direction to the photosphere. 

For economy space, we show in the third panel of Figure\ref{evolution_of_ionization_states} only the density profile of singly ionized Helium (HeII) and the doubly ionized Helium (HeIII). In this case, HeII clearly shows a peak which covers the height $z\approx$ between 1.5 Mm and 3 Mm. The peak moves with the time from the transition region to the photosphere and the maximum remains almost constant.The HeIII shows an increase more concentrated at altitudes that corresponds with the transition region. Due to the altitude at the beginning of the transition region drops against time steps, the peak of the HeIII also drops with the time.

Finally, in the fourth panel of Figure \ref{evolution_of_ionization_states}, we show the negative ionized Hydrogen density (H-) and the electronic density $n_e$. In this case, the negative ionized Hydrogen density H- shows a peak that moves with the time between $z\approx$ 2 Mm until 1.5 Mm. The peak shows an increase in density against the time step.
However, H- remains almost constant to altitudes lower than 1.5 Mm. The profile of $n_e$ is very close to the classical hydrostatic models. The differences of densities in each time steps are less than half of one order of magnitude compared with the initial condition. For the case of $n_e$ we found a similar wave observed in HI and HII. In the following paragraphs, we show the characterization of the observed wave.

\begin{figure*}
\centering
\includegraphics[width=8.0cm,height=5.0cm]{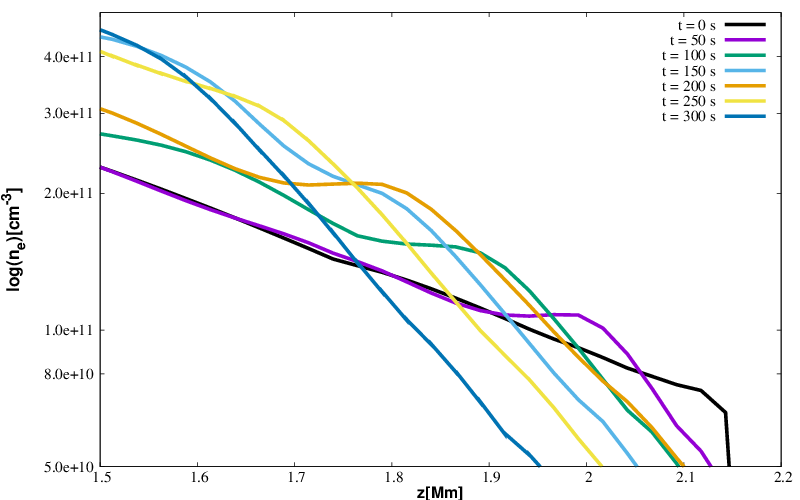}
\includegraphics[width=7.8cm,height=5.0cm]{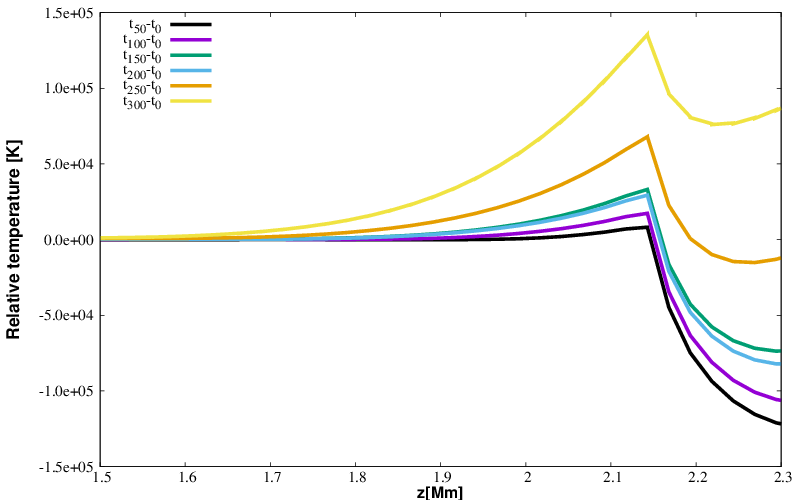}
\includegraphics[width=8.0cm,height=5.0cm]{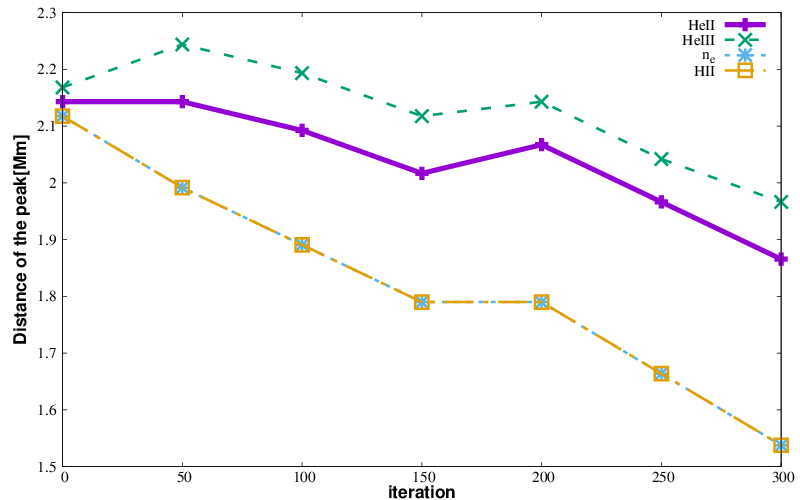}
\includegraphics[width=8.0cm,height=5.0cm]{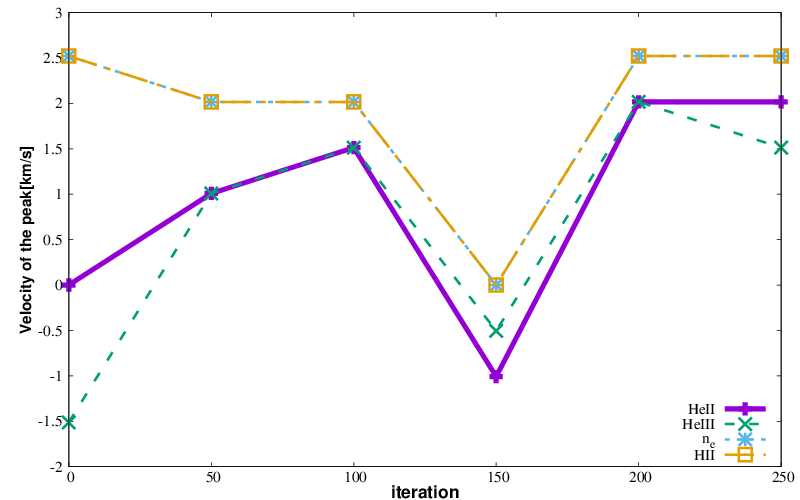}
\caption{(Top-left) Zoomed view of the logarithm of electronic density $n_{e}$ (cm$^{-3}$) for various times.  
(Top-right) Relative of the logarithm of temperature in K as function of $z$. (Bottom-left) Plot of the maximum distance of the peak as a function of time for HeII, HeIII, $n_e$ and HII. (Bottom-right) Velocity of the maximum of the peak as a function of the iterations for HeII, HeIII, $n_e$ and HII.}
\label{relative_H_temp}
\end{figure*}

{\it Wave in the electronic density profile.} In the profile of electronic density $n_e$ of Figure \ref{evolution_of_ionization_states} we can see oscillatory motions moving from the transition region to the photosphere along $z$, these features are also see in the profile of HI and HII. The perturbations are mainly seen between 1.5 and 2.1 Mm as show in the top-left of Figure \ref{relative_H_temp}, which represents a zoom of the electronic density $n_e$. We wonder if these perturbations are related with changes in density or temperature. To see more clearly the changes in temperature, we calculate the relative temperature defined as the difference between the temperature at later times with the temperature at initial time $T_{i}-T_{0}$, here $T_{i}$ is temperature at late times and $T_0$ denoted the temperature at initial time. For instance, on top-right of Figure \ref{relative_H_temp}, we show $T_{i}-T_{0}$ as function of $z$ in the region [1.5,2.3] Mm, in this case we can see that the difference in temperature is larger near to the transition region. This behavior of the temperature represents a process of thermalization, which in turns can affect the dynamic of electronic density. 

To characterize the perturbations, we obtain a time-distance diagram defined in terms of the position of the main peak and the iterations in time. We show the time-distance diagram on the bottom-left of Figure \ref{relative_H_temp}, where we can identify that the behavior of $n_e$ and HII are perfectly correlated, which means that 
the wave observed in the profiles of $n_e$ corresponds directly with changes in HII. Finally, we use the time-distance diagram to estimate the velocity of the wave, in this case we can see velocities around 2.5 km/s. However we can observe a deceleration between iterations 150 and 200, which means that thermalization process slow downs in that period of time.

\begin{figure}
\centering
\includegraphics[width=8.0cm,height=5cm]{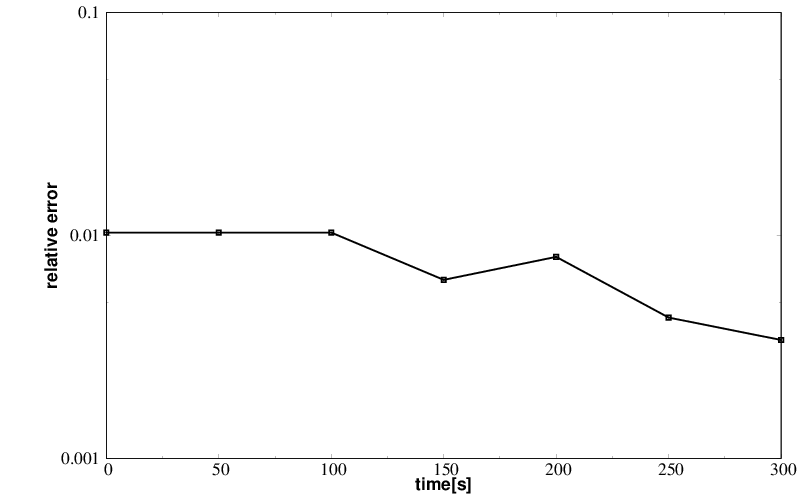}
\caption{Relative error $|1-\frac{\rho}{N}|$ as a function of time.}
\label{relative_error}
\end{figure}

{\it Convergence}. To test the convergence of our model, we estimate the conservation of total mass density of the plasma, i.e., we calculate the relative error between the total mass density $\rho$ obtained by CAFE and the total density N obtained with PakalMPI. In PakalMPI, we define N = HI + HII + HeI + HeII + HeIII, therefore we define the relative error as $|1-\frac{\rho}{N}|$. In Figure \ref{relative_error}, we show the relative error as a function of time for a set of points with an interval of 50 s between them. We can see that the relative error is of the order $10^{-2}$ and decreases with time, which indicates that the total mass density of the plasma is conserved.  

\section{Conclusions and final comments}
\label{conclusions}

We have developed a new numerical model capable to solve the MHD equations and the multispecies, whose ionization state densities are calculated by the NLTE approximation for HI, HII, H- and $n_e$ and the other species are computed by classical LTE approximation. 

The test of our code show the stability of the very complex process to obtain the densities for the species that compose the solar chromosphere. We prove that the numerical dispersion of our approximation is very low ($\approx 1\%$), the integration time of the MHD equations for the magnetohydrostatic system lasted approximately 5 days using 32 processors and the calculation of the ionization densities lasted a day and half using 32 processors.  

Our results show a hydrostatic chromospheric model, which is similar to previous results published in \cite{2016A&A...585A...4C}. The low region of the chromosphere remains unalterable but the transition region presents fast changes in temperature and density. The high difference in temperature between the corona and the chromosphere can not be sustained with the initial conditions of our model. We observe a fast thermalization in the high region of the chromosphere. At altitudes between 1.5 and 2.5 Mm over the photosphere the gradient of temperature increase with the time, producing a high rate of ionization at these altitudes. However, between 0 and 1.5 Mm the chromosphere is very stable for at least 300 s. 

The wave observed in the electronic density profiles are similar to the observed in the time-sequence of the intensities of submillimeter and millimeter continua in the solar chromosphere \citep{Loukitcheva_et_al_2004}. In such paper, the authors considered a collection of submillimeter and millimeter wave observed brightness temperatures $T_{b}$ of the quiet Sun and compared it with brightness temperatures computed from the standard static models of Fontella, Avrett and Loeser (FAL) \citep{Fontella_et_al_1993} and the dynamic simulations of Carlsson \& Stein \citep{Carlsson&Stein_1992,Carlsson&Stein_1995,Carlsson&Stein_1997,Carlsson&Stein_2002}. In particular, the dynamic simulation is based on a initial atmosphere in radiative equilibrium disturbed by waves driven trough the atmosphere by a subphotospheric piston. These waves propagate and increase in amplitude and form shocks above a height of 1 Mm. In contrast to the previous model, in our simulation we do not perturb the equilibrium state, instead, we let to evolve the system and find the appearance of disturbances in the electronic density that resemble oscillatory behavior. In the case of our model, this wave can be explained as the result of local changes in the rate of ionization of HII in NLTE. The thermalization process in the top of the chromosphere could play an important rol in local changes of electronic density at these altitudes.

Finally, the model presented in this paper represents the initial step to develop a more general model capable to solve the full MHD equations coupled with the ionization state equations in NLTE and the calculation of the emission by solving the radiative transfer equation.  

\acknowledgments
This research is partly supported by CONACyT CIENCIA B\'ASICA 254497 and CONACyT - REPOSITORIO INSTITUCIONAL DE CLIMA ESPACIAL 268273. The simulations were carried out in the facilities of the Center of Supercomputing of Space Weather (CESCOM) part of the National Laboratory of Space
Weather (LANCE) in Mexico. J.J.G.-A gratefully acknowledges DGAPA postdoctoral grant to Universidad Nacional Aut\'onoma de M\'exico (UNAM). Visualization of the 3D simulation data was done with the use of Mayavi \citep{Ramachandran&Varoquaux_2011}.

\end{document}